\documentclass[amsmath,amssymb,aps,nofootinbib,showpacs,eqsecnum]{revtex4-2}

\usepackage{amsmath}
\usepackage{bm}
\usepackage{cancel}
\usepackage{cases}
\usepackage{CJKutf8}
\usepackage{color}
\usepackage{comment}
\usepackage{dcolumn}
\usepackage{graphicx}
\usepackage{here}
\usepackage{mathtools}
\usepackage{multirow}
\usepackage{slashed}
\usepackage{subfigure}
\usepackage{textcase}
\usepackage{ytableau}
\usepackage[colorlinks=true,citecolor=blue,linkcolor=blue,urlcolor=blue]{hyperref}
\usepackage[mathlines]{lineno}

\begin{document}

\begin{flushright}
KOBE-COSMO-25-08, KUNS-3048
\end{flushright} 

\title{
Graviton-photon conversion in stochastic magnetic fields
}

\author{Wataru Chiba$^{1}$}
\thanks{{\color{blue}245s114s@stu.kobe-u.ac.jp}}

\author{Ryusuke Jinno$^{1}$}
\thanks{{\color{blue}jinno@phys.sci.kobe-u.ac.jp}}

\author{Kimihiro Nomura$^{2}$}
\thanks{{\color{blue}k.nomura@tap.scphys.kyoto-u.ac.jp}}

\affiliation{$^{1}$
Department of Physics, Graduate School of Science, Kobe University, 1-1 Rokkodai, Kobe, Hyogo 657-8501, Japan
}

\affiliation{$^{2}$
Department of Physics, Kyoto University, Kyoto 606-8502, Japan
}

\date{\today}

\begin{abstract}
We study graviton-photon conversion in the presence of stochastic magnetic fields.
Assuming Gaussian magnetic fields that may possess nontrivial helicity, and unpolarized gravitational waves (GWs) as the initial state, we obtain expressions for the intensity and linear/circular polarizations of GWs after propagation over a finite distance.
We calculate both the expectation values and variances of these observables, and find their nontrivial dependence on the typical correlation length of the magnetic field, the propagation distance, and the photon plasma mass.
Our analysis reveals that an observationally favorable frequency range with narrower variance can emerge for the intensity, while a peak structure appears in the expectation value of the circular polarization when the magnetic field has nonzero helicity.
We also identify a consistency relation between the GW intensity and circular polarization.
\end{abstract}

\maketitle


\section{Introduction}
\label{sec:introduction}

Magnetic fields are ubiquitous in the universe, appearing across a vast range of scales—from planets and stars~\cite{Donati:2009xd} to galaxies and galaxy clusters~\cite{Bonafede:2010xg}. 
Remarkably, it has even been suggested from blazar observations that magnetic fields exist on cosmological scales such as in the intergalactic medium and cosmic voids~\cite{Neronov:2010gir, Taylor:2011bn}.
The origin of these large-scale magnetic fields, hereafter referred to as cosmic magnetic fields, remains one of the outstanding open problems in modern cosmology. 
The presence of the cosmic magnetic fields has also been explored by various observational probes, including Faraday rotation measurements~\cite{Kronberg:1993vk} and CMB analysis~\cite{Planck:2015zrl}.
Although their precise generation mechanism is yet to be established, they suggest a primordial or early-universe origin, considering the uncertainty of the seeds for the dynamo mechanism (see, e.g., Ref.~\cite{Durrer:2013pga}).
Several theoretical scenarios have been proposed to account for their generation, among which inflationary magnetogenesis and phase transition-driven mechanisms represent two typical magnetogenesis scenarios in the early universe~\cite{Subramanian:2015lua,Vachaspati:2020blt}.

Recent advances in gravitational wave (GW) astronomy have opened a new observational window into the early universe.
Since the first direct detection by LIGO in 2015~\cite{LIGOScientific:2016aoc}, numerous compact binary coalescences have been observed~\cite{LIGOScientific:2016sjg,LIGOScientific:2017vwq}, and attention has increasingly turned to the detection of a stochastic GW background of primordial origin. 
Such a background, possibly generated from cosmological dynamics such as inflation~\cite{Starobinsky:1979ty}, preheating~\cite{Kofman:1994rk}, topological defects~\cite{Vilenkin:1981bx,Vachaspati:1984gt}, cosmological phase transitions~\cite{Witten:1984rs,Hogan:1986dsh}, as well as from the high-temperature thermal bath~\cite{Ghiglieri:2015nfa,Ghiglieri:2020mhm,Ringwald:2020ist,Vagnozzi:2022qmc}, carries invaluable information about physics at energy scales far beyond the reach of terrestrial experiments.

In the presence of cosmic magnetic fields, GWs can undergo nontrivial interactions with the background field.
One prominent example is the Gertsenshtein effect (graviton-photon conversion), whereby GWs propagating in a magnetic field can convert into electromagnetic waves, and vice versa~\cite{gertsenshtein1962wave,PhysRevD.37.1237}. 
This effect gains renewed significance in cosmology, where magnetic fields of primordial origin may extend over megaparsec scales and interact with GWs originating from astrophysical phenomena or from  the early universe. 
This interaction not only provides a potential indirect probe of cosmic magnetic fields but also opens the possibility of modifications to the polarization state or spectral characteristics of the GW background, which could be observable with future high-sensitivity detectors~\cite{Fujita:2020rdx,Kushwaha:2025mia}\footnote{
Graviton-photon conversion has also been studied in the context of strong magnetic fields around compact objects~\cite{Ito:2023fcr,Ito:2023nkq,Matsuo:2025blj} or that of cavity experiments~\cite{Berlin:2021txa,Gatti:2024mde}.
}.

The formulation of graviton-photon conversion in the presence of a background magnetic field has often been done analytically assuming a constant magnetic field~\cite{Boccaletti:1970pxw,osti_4377804,Ejlli:2020fpt}.
To account for the more realistic scenario of cosmic magnetic fields, which are expected to be stochastic in nature, other developments have incorporated both ``domain-like'' models~\cite{Cillis:1996qy,Pshirkov:2009sf,Marsh:2021ajy, Nomura:2024cku} and perturbative approaches~\cite{Chen:2013gva,Fujita:2020rdx}.
In particular, the stochastic perturbative formalism models the magnetic field as a Gaussian random field characterized by a power spectrum~\cite{Domcke:2020yzq,Addazi:2024kbq,Li:2025eoo}.

In this work, we adopt the stochastic perturbative approach and investigate 
the conversion effect that occurs while GWs propagate through random magnetic fields.
We extend the analysis by including helical components in the magnetic power spectrum, while previous studies have primarily considered non-helical magnetic spectrum and focused on modifications to the intensity of GWs~\cite{Domcke:2020yzq,Addazi:2024kbq}.
Furthermore, we focus on the polarization properties of GWs, which are sensitive to parity-violating effects induced by helical magnetic fields.
Our analysis reveals distinctive features in the GW polarizations that arise from both the helicity of the background field and the finite propagation distance, offering a potential probe of parity violation in the universe.For this reason, we concentrate on the conversion effect on GWs emitted from certain sources.
Since the signals of electromagnetic waves converted from GWs are likely to be weak and obscured by other electromagnetic radiation, we do not address them in this study.
Also, we do not consider conversion from electromagnetic waves into GWs.

More specifically, in this study we provide a systematic calculation of the polarization properties of GWs traversing stochastic magnetic fields that potentially have nontrivial helicity. 
By performing a perturbative expansion and exploiting the statistical properties of the magnetic field, we analytically compute not only the expectation values of the Stokes parameters of GWs but also their variances. 
We reveal how stochastic magnetic fields imprint themselves onto GWs polarization and also derive consistency relations between the variances and expectation values of the Stokes parameters.
Our results uncover the emergence of nontrivial structures in the polarization pattern, whose characteristics depend sensitively on model parameters such as the correlation length of the magnetic field and the photon's effective mass induced by plasma.

The present paper is organized as follows.
In Sec.~\ref{sec:review}, we review the propagation of the graviton-photon system.
We derive basic equations describing the conversion of GWs into electromagnetic waves through a background magnetic field.
In Sec.~\ref{sec:random}, we derive formulas for the statistical quantities (expectation values and variances) for the observables in this system (Stokes parameters), assuming that the magnetic field is Gaussian and described by power spectra.
We also derive a consistency relation that holds independently of the power spectra.
In Sec.~\ref{sec:numerical} we investigate the numerical behavior of these statistical quantities as well as that of the convolution kernels and typical integrals.
Sec.~\ref{sec:conclusions} is devoted to discussion and conclusions.

\section{Review on graviton-photon system}
\label{sec:review}

\subsection{Action and equations of motion}

We first derive the Schr\"{o}dinger-like equation that describes the propagation of the graviton-photon system in a background magnetic field.
For details of the derivation, we refer the reader to Refs.~\cite{Masaki:2018eut, Ito:2023fcr, Addazi:2024kbq}.
We start from the Einstein-Hilbert and Maxwell action 
\begin{align}
    S=\int d^4x\sqrt{-g}~\mathcal{L}\,,
    \quad \mathcal{L}=\frac{M_{\rm P}^2}{2}R-\frac{1}{4}F_{\mu\nu}F^{\mu\nu}\,,
\end{align}
where $g$ is the determinant of the metric $g_{\mu\nu}$, $M_{\rm P} = 2.4\times10^{18}\,\mathrm{GeV}$ is the reduced Planck mass, $R$ is the Ricci scalar, $F_{\mu\nu}=\partial_\mu\mathcal{A}_\nu-\partial_\nu\mathcal{A}_\mu$ is the field strength tensor of the photon field $\mathcal{A}_\mu$.
From this action, the equations of motion are derived as
\begin{align}
\begin{cases}
    \nabla_\mu F^{\mu\nu}=0\,,\\[0.1cm]
    G_{\mu\nu}=\displaystyle\frac{1}{M_{\rm P}^2}\left(g^{\alpha\beta}F_{\alpha\mu}F_{\beta\nu}-\frac{1}{4}g_{\mu\nu}F_{\alpha\beta}F^{\alpha\beta}\right)
    \,,
\end{cases}
\end{align}
where $\nabla_\mu$ represents the covariant derivative and $G_{\mu\nu}$ is the Einstein tensor.
We consider a background magnetic field ${B}_i\equiv\epsilon_{ijk}\partial_j\bar{A}_k$, and GWs on Minkowski spacetime:
\begin{align}
    \begin{cases}
        \mathcal{A}_\mu=\bar{A}_\mu+A_\mu\,,\\[0.1cm]
        F_{\mu\nu}=\bar{F}_{\mu\nu}+f_{\mu\nu}\,,\\[0.1cm]
        g_{\mu\nu}=\displaystyle\eta_{\mu\nu}+\frac{2}{M_{\rm P}}h_{\mu\nu}\,,
    \end{cases}
\end{align}where $\bar{A}_\mu$ is the background photon field, $\bar{F}_{\mu\nu}$ is its field strength, and $\eta_{\mu\nu}$ is the background Minkowski metric.
We impose the transverse-traceless gauge condition for GWs,
\begin{align}
    h_{0\mu}=0\,,\quad\partial_jh_{ij}=0\,,\quad h_{ii}=0\,,
\end{align}
and also the radiation gauge condition for electromagnetic waves,
\begin{align}
    A_\mu=(0,A_i)\,,\quad\partial_i A_i=0\,.
\end{align}
We assume a situation where the graviton-photon system propagates in the presence of plasma particles, such as in external galactic space, and for this reason we include the effective plasma mass for photons,
\begin{align}
m_{\rm pl} &= \sqrt{ \frac{n_e e^2}{m_e} }\,,
\end{align}
where $n_e$ is the number density of 
electrons, $e^2 / (4\pi) \approx 1/137$ is the fine-structure constant, and $m_e$ is the electron mass.
In this paper, we treat the plasma density as spatially uniform.
The resulting equations of motion are then
\begin{align}
\begin{cases}
(\partial^2-m_{\rm pl}^2)A_i=\displaystyle\frac{2}{M_{\rm P}}\epsilon_{nml} {B}_l \, \partial_n h_{mi}\,,
\\[0.1cm]
\partial^2h_{ij}=\displaystyle\frac{1}{M_{\rm P}}(\epsilon_{nil} {B}_lf_{jn}+\epsilon_{njl} {B}_lf_{in})\,,
\end{cases}
\end{align}
where $\partial^2\equiv-\partial_t^2+\delta_{ij}\partial_i\partial_j$.

We take the propagation direction of the graviton-photon to be the $z$-direction.
We choose the following the basis vectors $e^{(n)}_i$ $(n=x,y,z)$ and expand the electromagnetic field and background magnetic field as
\begin{align}
\begin{cases}
    A_i = \displaystyle\sum_{n=x,y} e_i^{(n)} A_n(z) \, e^{-i\omega t} + \text{c.c.}
    = (A_x(z) e_i^{(x)} + A_y(z) e_i^{(y)}) \, e^{-i\omega t} + \text{c.c.}
    \,,\\[0.1cm]
    {B}_i = B_x e_i^{(x)} + B_y e_i^{(y)} + B_z e_i^{(z)}\,,
\end{cases}
\end{align}
where 
\begin{align}
    e_i^{(x)}=
    \begin{pmatrix}
        1\\
        0\\
        0\\
    \end{pmatrix}
    \,,\quad
     e_i^{(y)}=
    \begin{pmatrix}
        0\\
        1\\
        0\\
    \end{pmatrix}
    \,,\quad
    e_i^{(z)}=
    \begin{pmatrix}
        0\\
        0\\
        1
    \end{pmatrix}
    \,.
\end{align}
Similarly, we expand the GWs in plane waves using the basis tensors $\varepsilon^{(n)}_{ij}$ $(n=+,\times)$,
\begin{align}
    h_{ij}\equiv\sum_{n=+,\times}\varepsilon_{ij}^{(n)} h_n(z) e^{-i \omega t} + \text{c.c.}
    =(h_+(z) \varepsilon_{ij}^{(+)} + h_\times(z) \varepsilon_{ij}^{(\times)})\,e^{-i\omega t} + \text{c.c.}
    \,,
\end{align}
where
\begin{align}
    \varepsilon_{ij}^{(+)}
    \equiv \frac{1}{\sqrt{2}} (e_i^{(x)}e_j^{(x)}-e_i^{(y)} e_j^{(y)})
    =\frac{1}{\sqrt{2}}
    \begin{pmatrix}
        1&0&0\\
        0&-1&0\\
        0&0&0\\
    \end{pmatrix}
    \,,\quad
    \varepsilon_{ij}^{(\times)}
    \equiv \frac{1}{\sqrt{2}} (e_i^{(x)}e_j^{(y)}+e_i^{(y)}e_j^{(x)})
    =\frac{1}{\sqrt{2}}
    \begin{pmatrix}
        0&1&0\\
        1&0&0\\
        0&0&0\\
    \end{pmatrix}
    \,.
\end{align}
Since the photons and gravitons propagate in the $z$-direction, the spatial derivative can be written as $\partial_i = \partial_z e_i^{(z)} = i k_z e_i^{(z)}$ using the basis vector, where $k_z$ is the wavenumber in the $z$-direction. 
We obtain
\begin{align}
    ( \partial^2 - m_{\rm pl}^2 ) 
    \begin{pmatrix}
        A_x \\
        A_y 
    \end{pmatrix}
    &= \frac{\sqrt{2}}{M_{\rm P}} 
    \begin{pmatrix}
        B_y \partial_z h_+ - B_x \partial_z h_\times \\
        B_x \partial_z h_+ + B_y \partial_z h_\times 
    \end{pmatrix}
    \,,
    \\[0.1cm]
    \partial^2 
    \begin{pmatrix}
        h_+ \\
        h_\times
    \end{pmatrix}
    &= \frac{\sqrt{2}}{M_{\rm P}} 
    \begin{pmatrix}
        - B_y \partial_z A_x - B_x \partial_z A_y \\
        B_x \partial_z A_x - B_y \partial_z A_y
    \end{pmatrix}
    \,.
\end{align}
Adopting relativistic approximation, we write $\partial^2 = (\omega + i\partial_z) (\omega - i\partial_z)=(\omega + i\partial_z)(\omega+k_z) \approx 2 \omega(\omega + i\partial_z) $
\footnote{
The wavenumber of photons with plasma mass propagating in the $z$ direction can be written as $k_z = n \omega$, where the refractive index $n$ satisfies $n^{-2}=1+ ( m_{\mathrm{pl}} / k_z )^2$.
Under the relativistic condition $m_{\mathrm{pl}} / k_z \ll 1$, we have $n \approx 1$, so the linearization of $\partial^2$ is justified (see, e.g., Ref.~\cite{PhysRevD.37.1237}).
}.
Then we arrive at the Schr\"{o}dinger-like equation
\begin{align}
    i \partial_z \Psi(z) = \mathcal{M}(z) \Psi(z)\,,
    \label{eq:equation}
\end{align}
where the ``wave function'' $\Psi$ is a four-component field with two degrees of freedom for photons ($A_x, A_y$) and two degrees of freedom for gravitons ($h_+, h_\times$),
\begin{align}
    \Psi(z)\equiv
    \begin{pmatrix}
    A_x(z)\\
    A_y(z)\\
    h_+(z)\\
    h_\times(z)\\
  \end{pmatrix}
  \,.
\end{align}
On the other hand, the mixing matrix $\mathcal{M}$ corresponds to the ``Hamiltonian'' of the Schr\"{o}dinger-like equation
\begin{align}
  \mathcal{M}(z)
  \equiv
   \begin{pmatrix}
    \Pi_\gamma & 0 &\Pi_{My} & - \Pi_{Mx}\\
    0 & \Pi_{\gamma} & \Pi_{Mx} &\Pi_{My}\\
    -\Pi_{My} & - \Pi_{Mx} & \Pi_{g} & 0\\
    \Pi_{Mx} & - \Pi_{My} & 0 & \Pi_{g}\\
    \end{pmatrix}
    \,,
\end{align}
where the entries are given by
\begin{align}
\Pi_\gamma \equiv - \omega + \frac{m_{\rm pl}^2}{2\omega}\,,\quad
\Pi_g \equiv - \omega\,,\quad
\Pi_{Mx} \equiv \frac{i}{\sqrt{2} M_{\rm P}} B_x(z)\,,\quad
\Pi_{My} \equiv \frac{i}{\sqrt{2} M_{\rm P}} B_y(z)\,.
\end{align}
The diagonal elements of the mixing matrix correspond to the dispersion relations of the waves, where $\omega$ is the angular frequency of the field $\Psi$.
On the other hand, the off-diagonal elements represent the coupling between gravitational and electromagnetic waves through the background magnetic fields $B_x$ and $B_y$.
This coupling induces a conversion phenomenon during wave propagation. Consequently, if a background magnetic field is present, the propagation of GWs is accompanied by conversion into electromagnetic waves, and \textit{vice versa}.
For this reason, in the following, we use the term \textit{propagation} to refer to the propagation of waves accompanied by conversion.

In passing, we offer a few remarks on the modification of the photon dispersion relation.
At low frequencies, the photon's plasma mass receives a few corrections from the background magnetic field, known as the Cotton-Mouton effect or Faraday rotation (see, e.g., Refs.~\cite{Ejlli:2018hke,Ejlli:2018ucq}).
These effects induce a mass difference between the two photon polarization states, thereby leading to the generation of polarization.
On the other hand, at high frequencies, quantum electrodynamics (QED) corrections give rise to an effective photon mass. This is commonly referred to as the Euler-Heisenberg effect~\cite{Heisenberg:1936nmg}.
However, these corrections turn out to be higher order in the magnetic field in the observables we are interested in, since we consider the regime where the magnetic field is sufficiently weak and thus we take the leading order effect in the magnetic field into account.
Consequently, the two photon polarization states satisfy the same dispersion relation, acquiring an effective mass solely from a uniform plasma background.

\subsection{Born approximation}

We solve Eq.~\eqref{eq:equation} using first-order perturbation theory (i.e., Born approximation), assuming that the magnetic field is sufficiently weak. 
We first decompose the mixing matrix into the zeroth and first order parts as
\begin{align}
  \mathcal{M}(z)=
    \begin{pmatrix}
    \Pi_\gamma & 0 & 0& 0\\
    0 & \Pi_\gamma & 0 & 0\\
    0 & 0 & \Pi_g&0 \\
    0 & 0 & 0 & \Pi_g\\
    \end{pmatrix}
    +
   \begin{pmatrix}
    0 & 0 &\Pi_{My} & - \Pi_{Mx}\\
    0 & 0 & \Pi_{Mx} &\Pi_{My}\\
    -\Pi_{My} & - \Pi_{Mx} & 0 & 0\\
    \Pi_{Mx} & - \Pi_{My} & 0 & 0 \\
    \end{pmatrix}
\equiv\mathcal{M}_0+\mathcal{M}_1(z)\,.
\end{align}
This resembles a quantum mechanical problem where a time-dependent perturbation Hamiltonian is added to the unperturbed Hamiltonian.
We solve this system in terms of the transfer matrix $\mathcal{U}(z,0)$ defined by
\begin{align}
\Psi(z) = \mathcal{U}(z,0) \Psi(0)\,.
\end{align}
We expand the transfer matrix in terms of the perturbation order as
\begin{align}
\mathcal{U}(z,0) &= \mathcal{U}_0(z,0) + \mathcal{U}_1(z,0) + \cdots\,.
\end{align}
The zeroth order part $\mathcal{U}_0$ can be easily obtained as \begin{align}
\mathcal{U}_0(z,0) = e^{-i\mathcal{M}_0 z}\,.
\end{align}
Using this expression, the transfer matrix up to first-order in perturbation can be written as
\begin{align}
  \mathcal{U}(z,0) &= \mathcal{U}_0(z,0) \left[ I_{4\times4}-i \int_0^z ds~ \mathcal{U}_0^{\dagger}(s,0) \mathcal{M}_1(s) \mathcal{U}_0(s,0)\right]\nonumber\\
  &=\mathcal{U}_0(z,0) \left[ I_{4\times4} + \frac{1}{\sqrt{2}M_{\rm P}}\int_0^z ds~
  \begin{pmatrix}
    \textbf{0}_{2\times2}& e^{i\Pi(\omega)s}\textbf{B}^\top(s)\\
    - e^{-i \Pi(\omega)s}\textbf{B}(s)&\textbf{0}_{2\times2}\\
  \end{pmatrix}
  \right]\,,
  \label{eqUsol}
\end{align}
where $I_{4 \times 4}$ is the $4 \times 4$ identity matrix,
\begin{align}
\textbf{B}\equiv
    \begin{pmatrix}
    B_y & B_x \\
    -B_x & B_y\\
    \end{pmatrix}
    ~,
    \quad 
    \Pi(\omega)\equiv\Pi_\gamma-\Pi_g=\frac{m^2_{\rm pl}}{2\omega}\,,
\end{align}
and $\mathbf{B}^\top$ is the transpose of $\mathbf{B}$.

\subsection{Density matrix and observables}

We finally discuss the polarization states of the electromagnetic waves and GWs, and introduce Stokes parameters as observables.
It is convenient to define the density matrix as $\rho=\Psi\otimes\Psi^\dagger$, which is a quantity that describes the classical probabilistic mixed state of the system.
In this paper we are interested in how GWs without statistical polarization at the initial state get converted into electromagnetic waves during propagation through a magnetic field.
Thus we choose the initial state as
\begin{align}
\rho_{\rm ini} = \frac{1}{2}
\begin{pmatrix}
0&0&0&0\\
0&0&0&0\\
0&0&I_0+Q_0&U_0+iV_0\\
0&0&U_0+iV_0&I_0-Q_0\\
\end{pmatrix}= \frac{1}{2}
\begin{pmatrix}
0&0&0&0\\
0&0&0&0\\
0&0&1&0\\
0&0&0&1\\
\end{pmatrix}
\,,
\end{align}
where $I_0$ represents the intensity corresponding to the total amount of GWs, $Q_0$ and $U_0$ represent linear polarizations, and $V_0$ represents circular polarization of the GWs.
In what follows, we refer to the conditions $I_0=1$ and $Q_0=U_0=V_0=0$ as the \textit{unpolarized state}. Thus, the above state $\rho_{\mathrm{ini}}$ corresponds to a situation in which only unpolarized GWs are initially present.

Then the state after propagating a finite distance through a magnetic field can be calculated using the transfer matrix as
\begin{align}
  \rho(z) =\mathcal{U}(z,0) \rho_{\rm ini} \, \mathcal{U}^\dagger(z,0)
    =\frac{1}{2} I_{4 \times 4} - \frac{1}{2} \mathcal{U}(z,0)
\begin{pmatrix}
1&0&0&0\\
0&1&0&0\\
0&0&0&0\\
0&0&0&0\\
\end{pmatrix}
\mathcal{U}^\dagger(z,0)
\,.
\label{eqrho}
\end{align}
Here, the $2 \times 2$ submatrix in the bottom right of the $4 \times 4$ density matrix $\rho$ corresponds to the GW state, and the statistical polarization states of GWs can be described by these four components.
This is commonly referred to as the Stokes parameters defined as~\cite{Seto:2008sr}
\begin{align}
    \begin{cases}
    I(\omega)\equiv |h_+|^2 + |h_\times|^2 = \rho_{33}+\rho_{44}\,,\\[0.1cm]
    Q(\omega)\equiv |h_+|^2 - |h_\times|^2 = \rho_{33}-\rho_{44}\,,\\[0.1cm]
    U(\omega)\equiv h_+ h_\times^* + h_\times h_+^* = \rho_{34}+\rho_{43}\,,\\[0.1cm]
    V(\omega)\equiv i ( h_+ h_\times^* - h_\times h_+^* ) = i (\rho_{34}-\rho_{43})\,.
  \end{cases}
\end{align}
Substituting the solution of $\mathcal{U}(z,0)$ obtained in Eq.~\eqref{eqUsol} into Eq.~\eqref{eqrho}, one finds the expressions for the Stokes parameters $I$ and $V$ as
\begin{align}
     I(\omega) &= 1-a(\omega) \,,
     \label{eq:Ia}\\
     V(\omega) &= -i \, b(\omega) \,,
     \label{eq:Vb}
\end{align}
where $a$ and $b$ are given by
\begin{align}
  a(\omega)&= \frac{1}{2M_{\rm P}^2} \int_0^z ds \int_0^z ds' e^{-i \Pi(\omega) (s-s')}
  [ B_x(s) B^*_x(s') + B_y(s) B^*_y(s') ]\nonumber\\
  &= \frac{1}{2M_{\rm P}^2} \int_0^z ds \int_0^z ds' e^{-i \Pi(\omega) (s-s')}
  [ B_+(s) B^*_+(s') + B_-(s) B^*_-(s') ]
  \,,
  \label{eq:a}\\
  b(\omega)&= \frac{1}{2M_{\rm P}^2} \int_0^z ds \int_0^z ds' e^{-i \Pi(\omega) (s-s')}
  [ B_x(s) B^*_y(s') - B_y(s) B^*_x(s') ]\nonumber\\
  &= \frac{-i}{2M_{\rm P}^2} \int_0^z ds \int_0^z ds' e^{-i \Pi(\omega) (s-s')}
  [ B_+(s) B^*_+(s') - B_-(s) B^*_-(s') ]
  \,.
  \label{eq:b}
\end{align}

Note that $a$ is real while $b$ is pure-imaginary.
In these expressions,
the magnetic fields are described in the orthogonal basis ($B_x,B_y$) in the first row, and in the helical basis ($B_+,B_-$) in the second row, respectively.
The transformation between the two is given by
\begin{align}
    \begin{pmatrix}
        B_x\\
        B_y
    \end{pmatrix}
    =\frac{1}{\sqrt{2}}
    \begin{pmatrix}
        1&1\\
        i&-i\\
    \end{pmatrix}
    \begin{pmatrix}
        B_+\\
        B_-
    \end{pmatrix}
    \,.
\end{align}

These are quantities of second order in the magnetic field, and they are written as the coordinate integral $\int ds$ of the product of the magnetic field $B_i(s)~(i = x, y \,\mathrm{or}\,+, -)$ and the wave $e^{- i\Pi(\omega)s}$ with frequency $\Pi(\omega) \equiv \Pi_\gamma - \Pi_g = m_{\rm pl}^2 / 2 \omega$ characterizing the graviton-photon system.
On the other hand, one finds that the Stokes parameters $Q$ and $U$ identically vanish.

The  parameter $I$ represents the total amount of GWs after propagating a distance $d$.
In other words, $I$ corresponds to the probability that GWs survive conversion in the background magnetic field.
Hence, $a$ can be understood as the conversion probability of the GWs to the electromagnetic waves during propagation.
The parameters $Q$ and $U$ represent the linear polarization of GWs remaining after conversion over a propagation distance $d$, and the result $Q = U = 0$ shows that GWs in an unpolarized initial state do not exhibit linear polarization regardless of the magnetic field they pass through before converting into electromagnetic waves.

The parameter $V$ represents the circular polarization of GWs remaining after conversion over a propagation distance $d$.
The expression \eqref{eq:b} means that GWs in an unpolarized initial state exhibit circular polarization only when they pass through a helical magnetic field before converting into electromagnetic waves.
This property is also reflected in the final expression for the circular polarization in Eq.~\eqref{eq:Stokes_statistical}, in which the expectation value of $V$ depends only on $\beta$ determined from the helical part of the magnetic field power spectrum as seen from Eq.~\eqref{eq:abcd}.

\section{Propagation in stochastic magnetic fields}
\label{sec:random}

In this section we discuss statistical properties of the magnetic field, and express the expectation value and variance of the observables (Stokes parameters) in terms of its power spectra.

\subsection{Power spectra of the magnetic field and convolution kernels}
\label{subsec:abcd}

In this paper we assume that the magnetic field is stochastically realized with a homogeneous, isotropic and Gaussian distribution.
Because of the Gaussian nature, the statistical properties of the magnetic field is solely described by power spectra.
Statistically homogeneous, isotropic, and Gaussian magnetic fields are characterized by the two-point function~\cite{Durrer:2013pga,Brandenburg:2018ptt},
\begin{align}
  \left< B_i(\vec{x})B_j(\vec{x}')\right>=\int  \frac{d^3k}{(2\pi)^3}e^{i\vec{k}\cdot(\vec{x}-\vec{x}')}\Big[(\delta_{ij}-\hat{k}_i\hat{k}_j )P_B(k)-i\epsilon_{ijm}\hat{k}_mP_{aB}(k)\Big]\,,
  \label{2ptfunc}
\end{align}
where $k\equiv\sqrt{k_x^2+k_y^2+k_z^2}$ and $\hat{k}_i\equiv k_i/k$.
The bracket $\left< \cdots \right>$ denotes an ensemble average over infinitely many realizations of the stochastic magnetic field.
$P_B(k)$ and $P_{aB}(k)$ are the symmetric and antisymmetric part of the magnetic field spectrum, respectively.
The combination $(\delta_{ij} - \hat{k}_i \hat{k}_j)$ in Eq.~\eqref{2ptfunc} reflects the property that the magnetic field is divergence-free.
The factor $i$ in the second term is put so that $P_{aB}(k)$ takes real values as well as $P_B(k)$.
When $P_{aB}(k) \neq 0$, the magnetic field is referred to as helical.
The maximal helicity is realized when $|P_{aB}(k)| = P_B(k)$.
In the calculations below, we introduce the following notations for brevity,
\begin{align}
\int_k\equiv\int \frac{d^3k}{(2\pi)^3}\,,
\quad
\int_s\equiv\int_0^dds\,.
\end{align}
In the latter we consider propagation over a distance $d$.
Since the Stokes parameters are expressed in terms of the magnetic field, we calculate the expectation value and the variance of the Stokes parameters with respect to the ensemble of the magnetic field.

The expectation value $\mathrm{Exp}[O]$ and variance $\mathrm{Var}[O]$ of an observable $O$ with respect to the ensemble of the magnetic field are written in terms of the bracket $\left< \cdots \right>$ as
\begin{align}
    \mathrm{Exp}[O]\equiv\left< O \right>\,,
    \quad \mathrm{Var}[O]\equiv\left<(O-\left< O \right>)^2\right>=\left< O^2 \right>-\left< O \right>^2
    \,.
\end{align}
The expectation value of the Stokes parameters requires calculation of two-point functions of the magnetic field, while the variance involves calculation of four-point functions.
However, because of the Gaussian assumption on the magnetic field, four-point functions can be written as the product of two-point functions (Wick contractions), and thus the calculation essentially reduces to that of two-point functions.
As shown in Appendix \ref{app:Wick}, there are essentially only four patterns that appear in the two-point functions
\begin{align}
  \begin{cases}
    \alpha(\omega)\equiv\displaystyle\frac{1}{2 M_{\rm P}^2}  \int_k\mathcal{I}(k_z;\omega)\frac{1+\hat{k}^2_z}{2}P_{B}(k)\,,\\[0.1cm]
    \beta(\omega)\equiv\displaystyle\frac{1}{2 M_{\rm P}^2} \int_k\mathcal{I}(k_z;\omega)\hat{k}_z P_{aB}(k)\,,\\[0.1cm]
    \gamma(\omega)\equiv\displaystyle\frac{1}{2 M_{\rm P}^2} \int_k\mathcal{J}(k_z;\omega)\frac{1+\hat{k}_z^2}{2}P_{B}(k)\,,\\[0.1cm]
    \delta(\omega)\equiv\displaystyle\frac{1}{2 M_{\rm P}^2} \int_k\mathcal{J}(k_z;\omega)\hat{k}_z P_{aB}(k)=0\,.
  \end{cases}
    \label{eq:abcd0}
\end{align}
The observables related to GWs or electromagnetic waves can be described only by these integrals, as we see below.
Note that all these integrals are functions of the GW angular frequency $\omega$.
The integral kernels $\mathcal{I}$ and $\mathcal{J}$ are defined as
\begin{align}
  \mathcal{I}(k_z;\omega) &\equiv \int_{s,s'}e^{-i(\Pi(\omega)-k_z)(s-s')}=
  \bigg|\int_{s}e^{-i(\Pi(\omega)-k_z)s} \bigg|^2 \,,\\
  \mathcal{J}(k_z;\omega)
  &\equiv e^{i\Pi(\omega) d}\int_se^{-i(\Pi(\omega)  - k_z)s}\int_{s'}e^{-i(\Pi(\omega) + k_z)s'}\,.
\end{align}
The kernel $\mathcal{J}$ is an even function with regard to $k_z$.
One of the typical integrals $\delta$ vanishes because of the symmetry of $\mathcal{J}$ with respect to the $k_z$ direction.
$\Pi(\omega)$ that characterizes the graviton-photon mass difference is contained within these integral kernels.
The kernels upon integration over $s$ reduce to
\begin{align}
  \mathcal{I}(k_z;\omega)
  &= \left[ \frac{2d \sin (\theta_- (k_z;\omega)/2)}{\theta_- (k_z;\omega)} \right]^2\,,\\
  \mathcal{J}(k_z;\omega)
  &= \left[ \frac{2d \sin (\theta_+ (k_z;\omega)/2)}{\theta_+ (k_z;\omega)} \right]\left[ \frac{2d \sin (\theta_- (k_z;\omega)/2)}{\theta_- (k_z;\omega)} \right]\,,
\end{align}
where
\begin{align}
    \theta_\pm(k_z;\omega)\equiv(\Pi(\omega)\pm k_z)d\,.
\end{align}
The reason for the appearance of the four patterns $\alpha, \beta, \gamma$, and $\delta$ can be understood from (1) whether the magnetic field components in the two-point function are the same or different, and (2) whether the coordinate integrals give the kernel $\mathcal{I}$ or $\mathcal{J}$.
More specifically, ($\alpha$, $\beta$, $\gamma$, $\delta$) correspond to the case where the magnetic field components in the two-point function are (the same, different, the same, different) and the kernel type is ($\mathcal{I}$, $\mathcal{I}$, $\mathcal{J}$, $\mathcal{J}$), respectively.

The power spectra are functions of only the magnitude of the wavevector, and the integral kernels and coefficients are rotationally invariant in the $x$-$y$ plane of the wavevector space.
Thus, the four typical integrals discussed earlier possess rotational symmetry in this plane, allowing for a reduction in the number of integration variables. For practical purposes in numerical computations, we simplify the expressions into more manageable forms as
\begin{align}
  \begin{cases}
    \alpha= \displaystyle\frac{1}{8 \pi^2 M_{\rm P}^2} \int_0^\infty dk \,  P_B(k) \,  \mathcal{C}_{\alpha}(\Pi d,kd)\,,\\[0.1cm]
    \beta= \displaystyle\frac{1}{8 \pi^2 M_{\rm P}^2} \int_0^\infty dk \,  P_{aB}(k) \,  \mathcal{C}_{\beta}(\Pi d,kd)\,,\\[0.1cm]
    \gamma= \displaystyle\frac{1}{8 \pi^2 M_{\rm P}^2} \int_0^\infty dk \,  P_B(k) \,  \mathcal{C}_{\gamma}(\Pi d,kd)\,,\\[0.1cm]
    \delta= \displaystyle\frac{1}{8 \pi^2 M_{\rm P}^2} \int_0^\infty dk \,  P_{aB}(k) \,   \mathcal{C}_{\delta}(\Pi d,kd)\,.
  \end{cases}
  \label{eq:abcd}
\end{align}
Here, $\mathcal{C}_{i}$'s $(i=\alpha, \beta, \gamma, \delta)$ are the convolution kernels given by
\begin{align}
  \begin{cases}
       \mathcal{C}_{\alpha}(\Pi d,kd)\equiv\displaystyle(kd)^2\int_{-1}^1 dz~\left[ \frac{2\sin (\theta_-/2)}{\theta_-} \right]^2 \frac{1+z^2}{2}\,,\\[0.2cm]
       \mathcal{C}_{\beta}(\Pi d,kd)\equiv \displaystyle(kd)^2\int_{-1}^1dz~\left[ \frac{2 \sin (\theta_-/2)}{\theta_-} \right]^2 ~z\,,\\[0.2cm]
       \mathcal{C}_{\gamma}(\Pi d,kd)\equiv\displaystyle(kd)^2\int_{-1}^1dz~\left[ \frac{2\sin (\theta_+/2)}{\theta_+} \right]\left[ \frac{2 \sin (\theta_-/2)}{\theta_-} \right]\frac{1+z^2}{2}\,,\\[0.2cm]
       \mathcal{C}_{\delta}(\Pi d,kd)\equiv\displaystyle(kd)^2\int_{-1}^1dz~\left[ \frac{2\sin (\theta_+/2)}{\theta_+} \right]\left[ \frac{2 \sin (\theta_-/2)}{\theta_-} \right]~z=0\,,
    \end{cases}
    \label{eq:C}
\end{align}
with 
\begin{align}
    \theta_{\pm} = \Pi (\omega) d \pm kd \cdot z \,.
\end{align}
The integration over $z$ can be performed explicitly, but the resulting expressions are rather lengthy.
We summarize them in Appendix~\ref{app:C}.

\subsection{Expectation values and variances of the observables}
\label{subsec:Stokes}

The expectation value and variance of the intensity $I$ and circular polarization $V$ can be expressed in terms of the integrals in Eq.~\eqref{eq:abcd0} (or Eq.~\eqref{eq:abcd}).
Details are provided in Appendix~\ref{app:Wick}, and here we summarize the result,
\begin{align}
    \begin{cases}
     \mathrm{Exp}[1- I](\omega)=2\alpha\,,\\[0.1cm]
     \mathrm{Exp}[Q](\omega)=0\,,\\[0.1cm]
     \mathrm{Exp}[U](\omega)=0\,,\\[0.1cm]
     \mathrm{Exp}[V](\omega)= -2\beta\,,
  \end{cases}
  \quad 
  \begin{cases}
     \mathrm{Var}[1-I](\omega)=2(\alpha^2+\beta^2+\gamma^2-\delta^2)
     = 2(\alpha^2+\beta^2+\gamma^2) \,,\\[0.1cm]
     \mathrm{Var}[Q](\omega)=0\,,\\[0.1cm]
     \mathrm{Var}[U](\omega)=0\,,\\[0.1cm]
     \mathrm{Var}[V](\omega)=2(\alpha^2+\beta^2-\gamma^2+\delta^2)
     = 2(\alpha^2+\beta^2-\gamma^2)\,.
  \end{cases}
\label{eq:Stokes_statistical}
\end{align}
Since $\delta$ vanishes identically, one notices that four quantities $\mathrm{Exp}[1-I]$, $\mathrm{Exp}[V]$, $\mathrm{Var}[1-I]$, and $\mathrm{Var}[V]$ are written in terms of the three quantities $\alpha$, $\beta$, and $\gamma$.
Thus one may derive one nontrivial relation between these observables that holds regardless of the details of the power spectra
\begin{align}
  \mathrm{Var}[1-I]+\mathrm{Var}[V]=\mathrm{Exp}[1-I]^2+\mathrm{Exp}[V]^2\,.
  \label{eq:consistency}
\end{align}
We call it a consistency relation in the graviton-photon system.
If general relativity is correct and the magnetic field is described by the power spectra, then by statistically analyzing the intensity and polarization states of the initially unpolarized GWs propagating through random magnetic fields, the observables should reproduce this relation by measuring the GWs that remain after conversion within Born approximation.
This relation holds under the above assumptions, regardless of the details of the power spectra. 
However, if the initial GWs are nontrivially polarized, or if the Born approximation is invalid, this relation (\ref{eq:consistency}) does not necessarily hold.

The main results so far are the statistical expressions for the observables (\ref{eq:Stokes_statistical}) resulting from the propagation of initially unpolarized GWs propagating over a finite distance in a random magnetic field, the four types of integrals (\ref{eq:abcd})--(\ref{eq:C}), and the consistency relation (\ref{eq:consistency}).
It should be noted that all of these expressions are functions of the GW angular frequency $\omega$.
In the following section we analyze their properties numerically.

\section{Numerical analysis}
\label{sec:numerical}

In this section, we assume typical forms for $P_B (k)$ and $P_{aB} (k)$ and illustrate how the quantities in Sec.~\ref{sec:random} behave as functions of the angular frequency $\omega$ of GWs.
We first have a look at the behavior of the integral kernels, then integrals, and finally observables.

\subsection{\texorpdfstring{Kernels $\mathcal{C}_\alpha$, $\mathcal{C}_\beta$, and $\mathcal{C}_\gamma$}{Kernels Calpha, Cbeta, and Cgamma}}
\label{subsec:C}

\begin{figure}
\centering
\includegraphics[width=0.6\linewidth]{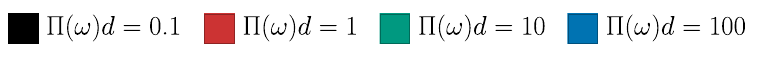}
\\[0.5cm]
\includegraphics[width=0.6\linewidth]{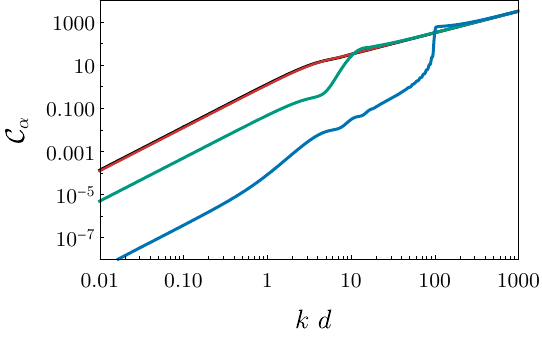}
\\[0.5cm]
\includegraphics[width=0.6\linewidth]{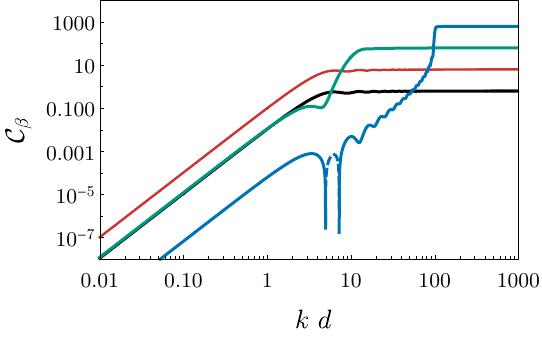}
\\[0.5cm]
\includegraphics[width=0.6\linewidth]{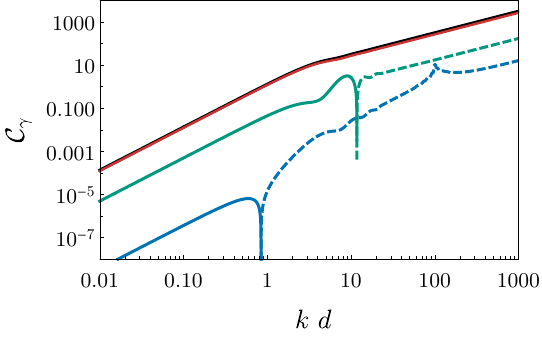}
\caption{\small
Convolution kernels $\mathcal{C}_\alpha$, $\mathcal{C}_\beta$, and $\mathcal{C}_\gamma$ with various values of $\Pi(\omega) d$ are shown as functions of $kd$. The black, red, green, and blue curves correspond to $\Pi(\omega) d = 0.1$, $1$, $10$, and $100$, respectively.
The dashed curves indicate that the kernel takes negative values.}
\label{fig:C}
\end{figure}

We show in Fig.~\ref{fig:C} the behavior of the dimensionless integral kernels $C_i~(i = \alpha, \beta, \gamma)$ as functions of $k d$ for various values of $\Pi (\omega) d$.
We do not plot $C_\delta$ since it vanishes identically.
Note that these kernels are the ones that should be convoluted with the power spectra of the magnetic field, and thus they are independent of the properties of the magnetic field.

From Fig.~\ref{fig:C}, several key properties of these kernels can be observed:
\begin{itemize}
\item 
In the case of $\Pi d \lesssim 1$, the kernels $\mathcal{C}_\alpha$ and $\mathcal{C}_\gamma$ approach the same functional form and become insensitive to the value of $\Pi d$. Meanwhile, the overall magnitude of the kernel $C_\beta$ scales proportionally with $\Pi d$. For all kernels, a transition in the power-law appears at $kd = \mathcal{O}(1)$. 
Their behaviors are $\mathcal{C}_\alpha (\approx \mathcal{C}_{\gamma}) \propto (kd)^2$ and $\mathcal{C}_\beta \propto (kd)^3$ for small $kd$, whereas $\mathcal{C}_\alpha (\approx \mathcal{C}_{\gamma}) \propto kd$ and $\mathcal{C}_\beta = \text{const}.$ for large $kd$.
\item
In the case of $\Pi d \gg 1$, each kernel appears to change its behavior around $kd = \mathcal{O}(1)$ and $kd \approx \Pi d$.
The kernels $\mathcal{C}_\alpha$ and $\mathcal{C}_\beta$ behave similarly to the case of $\Pi d \lesssim 1$ when $kd \gtrsim \Pi d$, while they are strongly suppressed when $kd \ll \Pi d$.
The kernel $\mathcal{C}_\gamma$ is suppressed over the entire range of $kd$ compared to the case of $\Pi d \lesssim 1$. 
These suppressions become stronger as $\Pi d$ increases.
When $kd \lesssim 1$, the kernels behave as $\mathcal{C}_\alpha (\approx \mathcal{C}_{\gamma}) \propto (kd)^2$ and $\mathcal{C}_\beta \propto (kd)^3$. When $1 \lesssim kd \lesssim \Pi d$, nontrivial modulations are observed. 
In particular, when $kd \gtrsim 1$, the kernel $\mathcal{C}_\gamma$ exhibits oscillatory changes in the sign depending on the value of $\Pi d$.
\end{itemize}

In Sec.~\ref{subsec:Stokes}, we saw that the combination $\alpha^2 - \gamma^2$ appears in the expression for the variance of $V$ (see Eq.~\eqref{eq:Stokes_statistical}).
Therefore, the observation that $\mathcal{C}_\alpha \approx \mathcal{C}_\gamma$ for $\Pi d \lesssim 1$ suggests that their cancellation leads to the nontrivial behavior of the variance in the bottom panel of Fig.~\ref{fig:V}, which we confirm later.

\subsection{\texorpdfstring{Integrals $\alpha$, $\beta$, and $\gamma$}{Integrals alpha, beta, and gamma}}
\label{subsec:abc}

\begin{figure}[htb]
\centering
\includegraphics[width=0.45\linewidth]{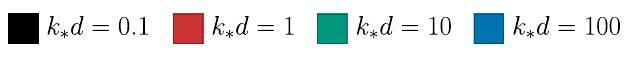}
\\
\includegraphics[width=0.45\linewidth]{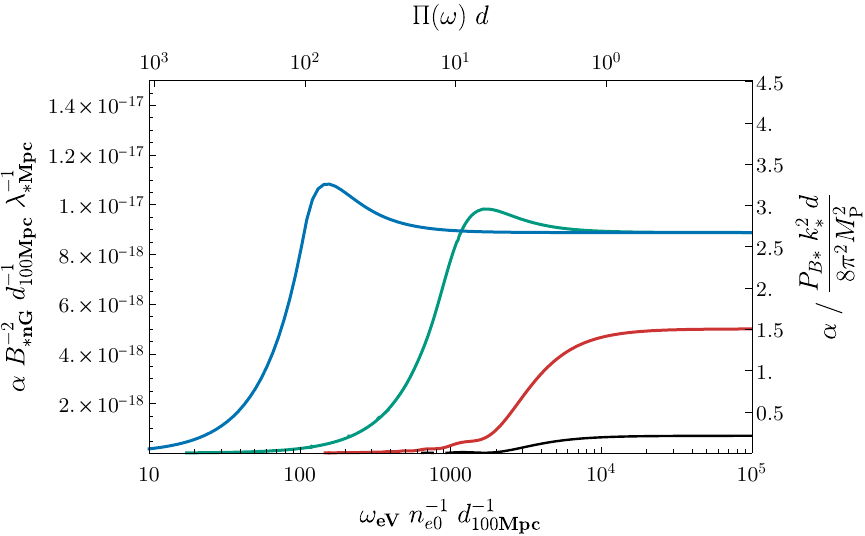}
\hskip 0.5cm
\includegraphics[width=0.45\linewidth]{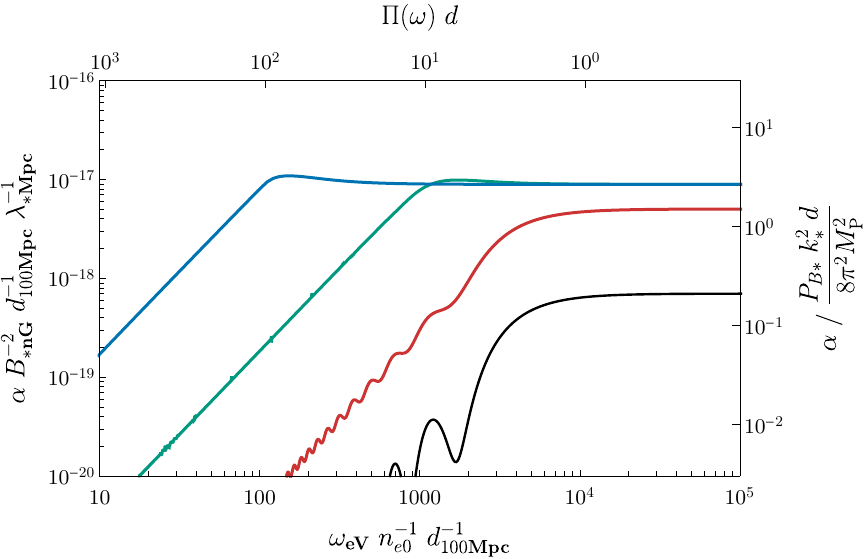}
\\
\includegraphics[width=0.45\linewidth]{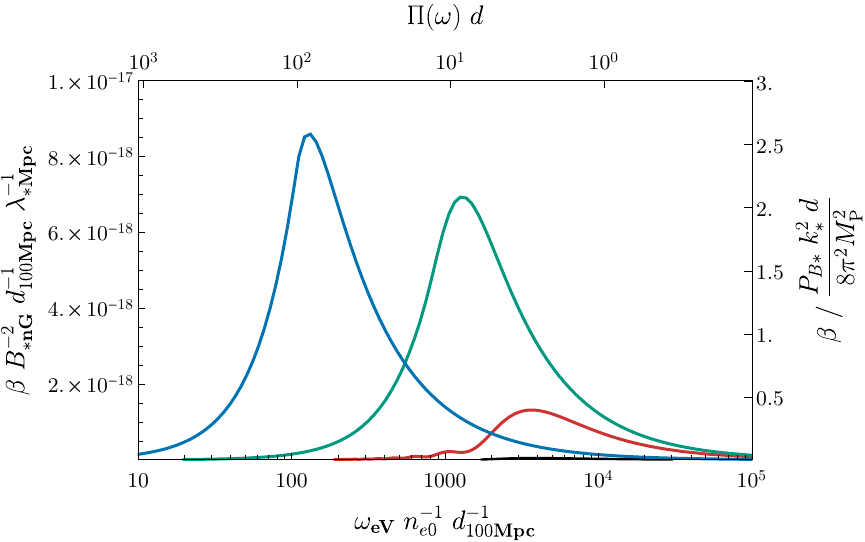}
\hskip 0.5cm
\includegraphics[width=0.45\linewidth]{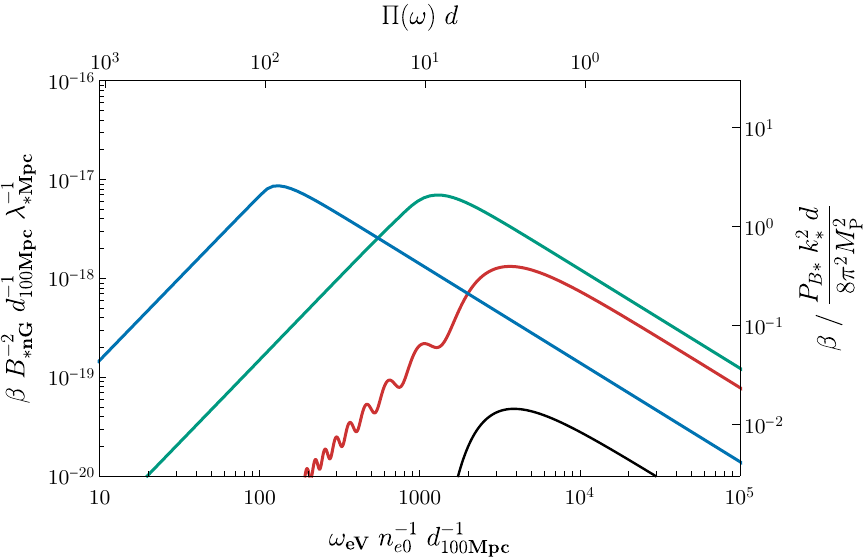}
\\
\includegraphics[width=0.45\linewidth]{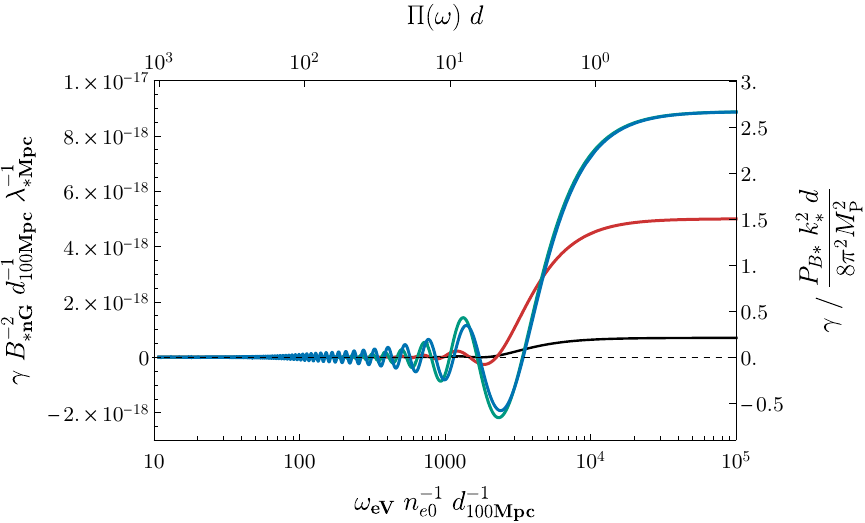}
\hskip 0.5cm
\includegraphics[width=0.45\linewidth]{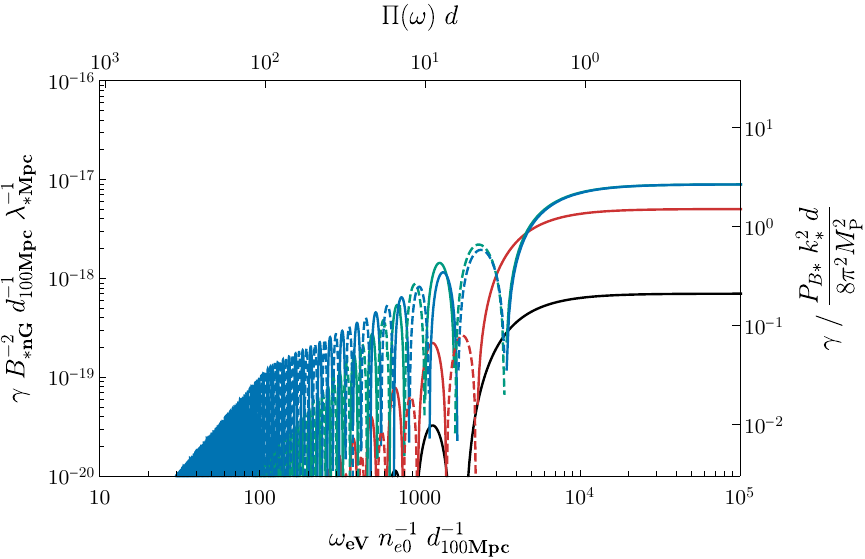}
\caption{
Integrals $\alpha$ (top), $\beta$ (middle), and $\gamma$ (bottom) with the magnetic field power spectra \eqref{eq:PB} and \eqref{eq:PaB} are shown as functions of GW angular frequency $\omega$.
The black, red, green, and blue curves correspond to the case of $k_* d = 0.1, 1, 10,$ and $100$, respectively.
In each panel, the upper horizontal axis is the dimensionless variable $\Pi(\omega) d$, and the lower horizontal axis is $\omega_{\text{eV}} \equiv \omega / \text{eV}$ normalized by $n_{e0} \equiv n_e / \text{m}^{-3}$ and $d_{100\text{Mpc}} \equiv d / (100\,\text{Mpc})$, where $n_e$ is the plasma density and $d$ is the propagation distance. 
The right vertical axis is the value of $\alpha$, $\beta$, or $\gamma$ normalized by $P_{B*} k_*^2 d / (8\pi^2 M_{\rm P}^2)$.
(For $\beta$, we set $P_{aB*} = P_{B*}$ assuming the maximally helical case.)
The left vertical axis is the value of $\alpha$, $\beta$, or $\gamma$ normalized by $B_{*\text{nG}}^2 \equiv (B_* / \text{nG})^2$, $d_{100\text{Mpc}}$, and $\lambda_{*\text{Mpc}} \equiv \lambda_{*} / \text{Mpc}$, where $B_{*}$ is the magnetic field strength at the peak scale $k_*$, and $\lambda_{*} \equiv 2\pi/k_*$ is the characteristic length.
In each row, the left and right panels show the same quantity: The vertical axis is linear in the left, but logarithmic in the right.
In the bottom-right panel, the dashed curves indicate that $\gamma$ is taking negative values.
}
\label{fig:abc}
\end{figure}

We next show in Fig.~\ref{fig:abc} the behavior of the integrals $\alpha$, $\beta$, and $\gamma$ from top to bottom, as functions of $\Pi (\omega) d$ for various values of $k_* d$.
For the power spectra of the magnetic field, we adopt the expressions in Ref.~\cite{Durrer:2013pga},
\begin{align}
    P_B(k) &= P_{B*} \times
    \begin{cases}
    \displaystyle\left(\frac{k}{k_*}\right)^{n_l}, \quad n_l = 2 \,, & (k \leq k_*)  \\
    \displaystyle\left(\frac{k}{k_*}\right)^{n_h}, \quad n_h = -\frac{11}{3} \,, & (k > k_*) 
    \end{cases}
    \label{eq:PB}
    \\
    P_{aB}(k) &= P_{aB*} \times
    \begin{cases}
    \displaystyle\left(\frac{k}{k_*}\right)^{n_{al}}, \quad n_{al} = 3 \,, & (k \leq k_*)  \\
    \displaystyle\left(\frac{k}{k_*}\right)^{n_{ah}}, \quad n_{ah} = - \frac{11}{3} \,, & (k > k_*) 
    \end{cases}
    \label{eq:PaB}
\end{align}
where $P_{B*}$ and $P_{aB*}$ are the peak amplitude of the symmetric and antisymmetric part of the power spectra, respectively (see Eq.~\eqref{2ptfunc}), and $k_*$ is the typical wavenumber of the magnetic field.
The exponents $n_l$ and $n_h$ ($n_{al}$ and $n_{ah}$) parametrize the scale dependence of $P_B$ ($P_{aB}$) for low and high wavenumbers, respectively.

In the following, we consider the maximally helical case $P_{B*} = P_{aB*}$ as the most illustrative example.
For non-helical case $P_{aB*} = 0$, see Appendix~\ref{app:non-helical}.
It is convenient to introduce the characteristic magnetic field strength at the peak scale $k_*$, denoted by $B_{*}$, as~\cite{Durrer:2013pga}
\begin{align}
    B_* \equiv \sqrt{\frac{k_*^3 P_{B*}}{\pi^2} }\,.
\end{align}
Additionally, we define the characteristic correlation length of the magnetic field as $\lambda_* \equiv 2\pi / k_*$. 
These quantities are used in the left vertical axes in Fig.~\ref{fig:abc}.

There are again several key properties in these plots:
\begin{itemize}
\item 
In the case of $k_* d \lesssim 1$, only $\Pi d = \mathcal{O} (1)$ appears as a characteristic scale for all of $\alpha$, $\beta$ and $\gamma$.
The overall amplitude of these integrals monotonically decreases as $k_* d$ decreases.
For the $\omega$-dependence, $\alpha$ and $\beta$ approach the same functional form for $\Pi d \gg 1$, while $\gamma$ is smaller in this frequency range.
The frequency dependence is estimated to be $\alpha = \beta \propto (\Pi d)^{- 5/3}$ while $\gamma \propto (\Pi d)^{- 2}$ in this range.
For $\Pi d \ll 1$, on the other hand, $\alpha$ and $\gamma$ approach the same form while $\beta$ is suppressed.
The frequency dependence is $\alpha = \gamma = \text{const}.$ while $\beta \propto \Pi d$.

\item
In the case of $k_* d \gg 1$, both $\Pi d = \mathcal{O} (1)$ and $\Pi d \approx k_* d$ appear as characteristic scales for $\gamma$, while only $\Pi d \approx k_* d$ appears for $\alpha$ and $\beta$.
Because of this, different regimes (1) $\Pi d \gg k_* d$ (2) $k_* d \gg \Pi d \gg 1$ and (3) $1 \gg \Pi d$ appear.
For low frequencies ($\Pi d \gg k_* d$), $\alpha$ and $\beta$ approach the same form $\alpha = \beta \propto (\Pi d)^{- 5/3}$ while $\gamma$ approaches $\gamma \propto (\Pi d)^{- 2}$.
For intermediate frequencies ($k_* d \gg \Pi d \gg 1$), $\alpha$ dominates over the other two integrals.
For high frequencies ($1 \gg \Pi d$), $\alpha$ and $\gamma$ approach the same functional form while $\beta$ is suppressed.
\end{itemize}

\subsection{Expectation values and variances of Stokes parameters}
\label{subsec:NumCalcStokes}

\begin{figure}
\centering
\includegraphics[width=0.45\linewidth]{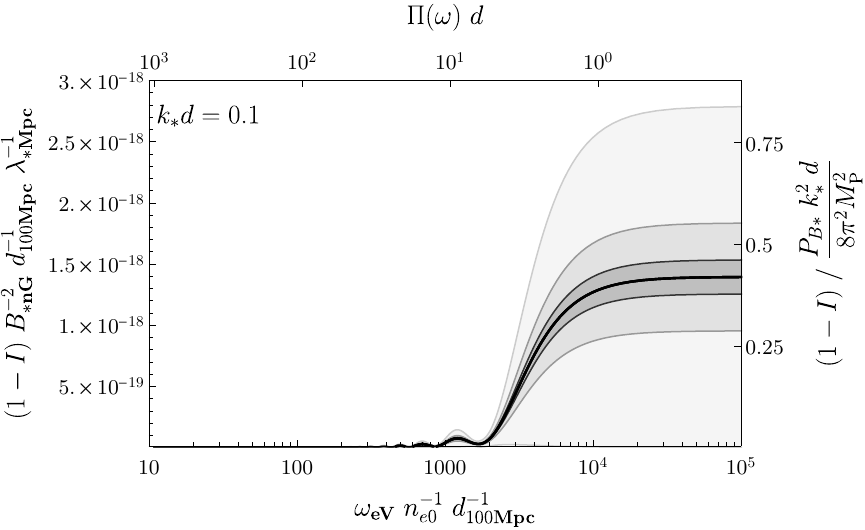}
\hskip 0.5cm
\includegraphics[width=0.45\linewidth]{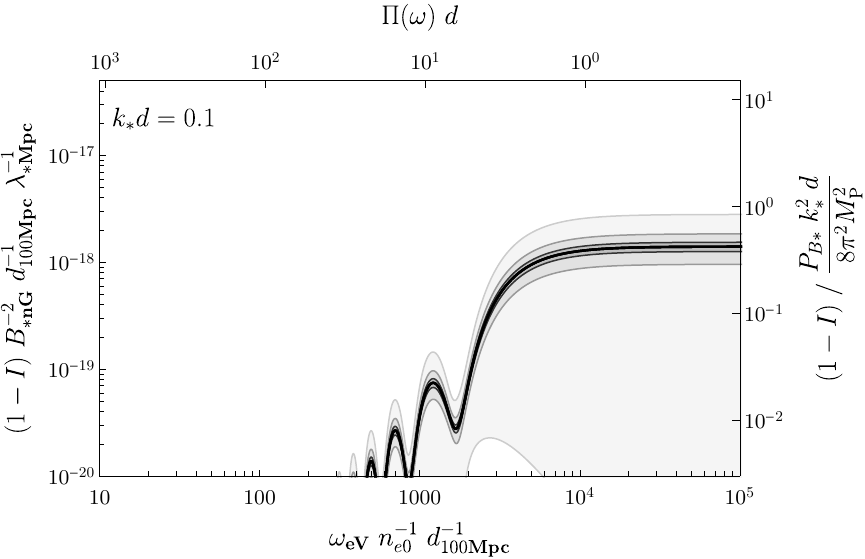}
\\
\includegraphics[width=0.45\linewidth]{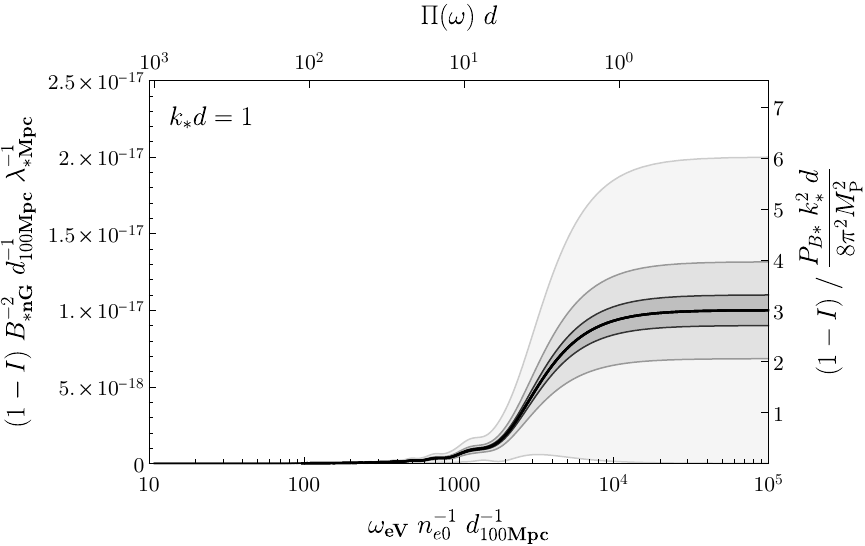}
\hskip 0.5cm
\includegraphics[width=0.45\linewidth]{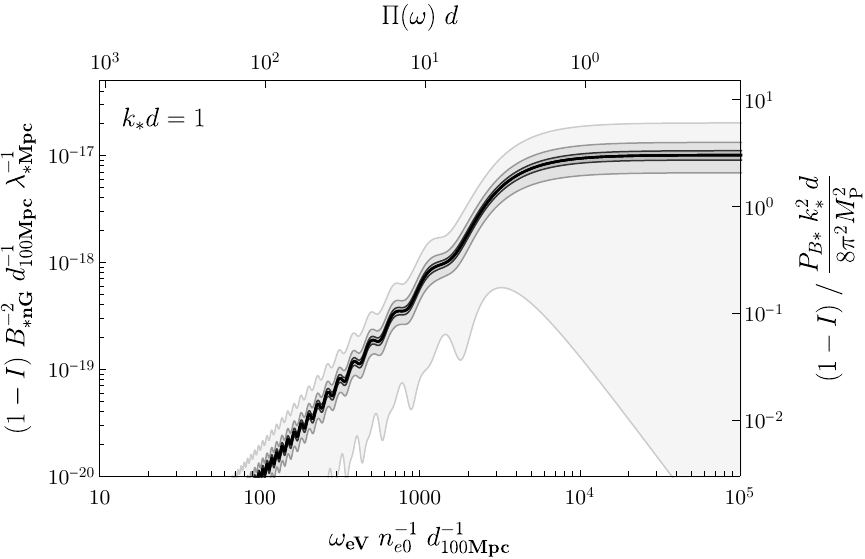}
\\
\includegraphics[width=0.45\linewidth]{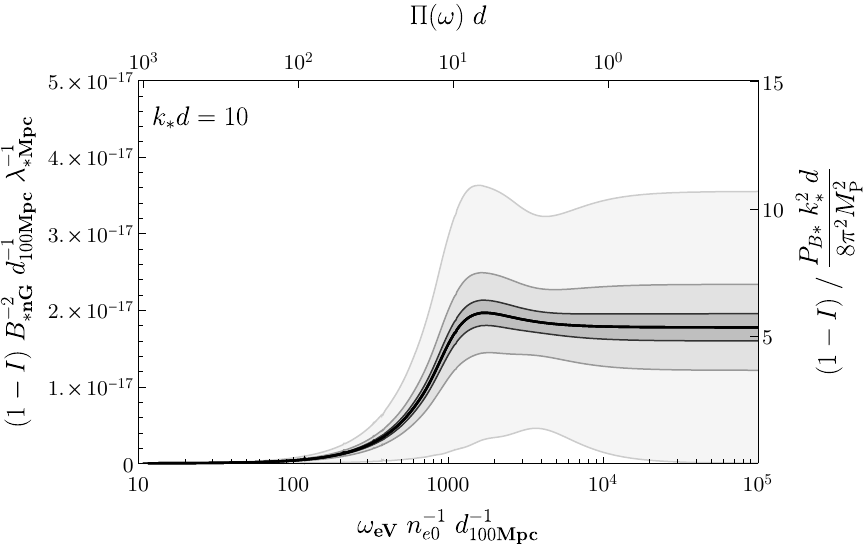}
\hskip 0.5cm
\includegraphics[width=0.45\linewidth]{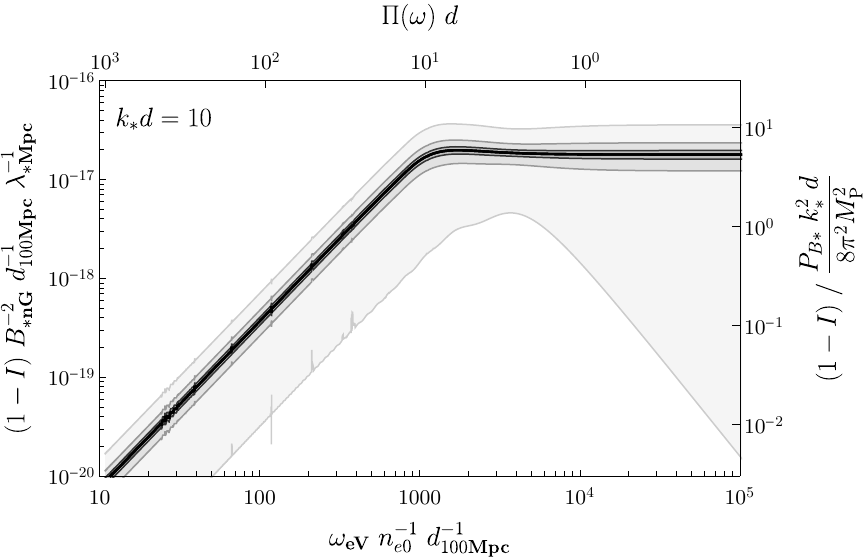}
\\
\includegraphics[width=0.45\linewidth]{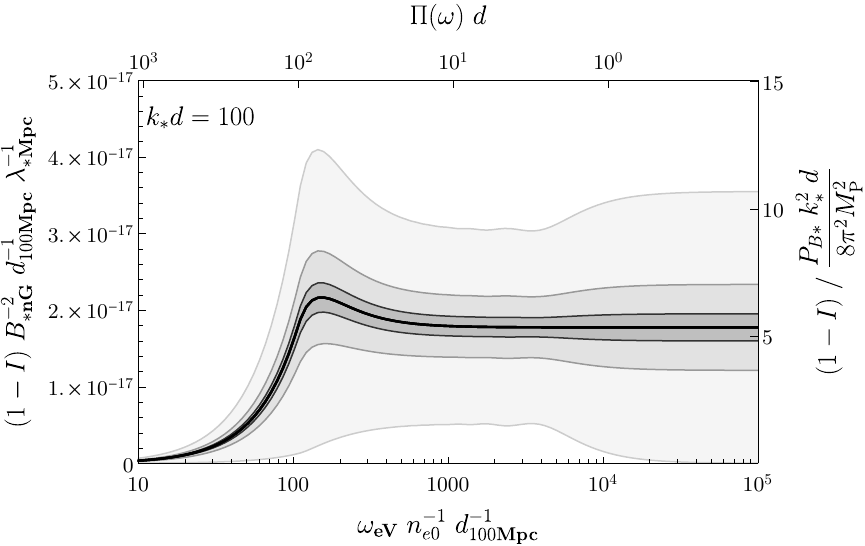}
\hskip 0.5cm
\includegraphics[width=0.45\linewidth]{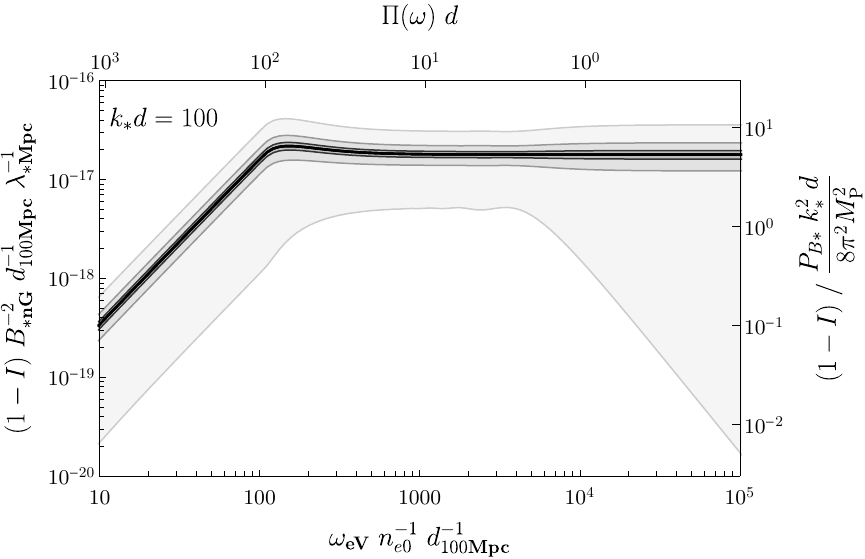}
\caption{
Conversion probability $P_{g\rightarrow\gamma} \equiv 1-I$ with the maximally helical ($P_{aB*} = P_{B*}$) magnetic field power spectra \eqref{eq:PB} and \eqref{eq:PaB} is shown as a function of GW angular frequency $\omega$ for various values of $k_* d$.
From top to bottom, the cases $k_* d = 0.1$, $1$, $10$, and $100$ are shown. 
In each panel, the black line represents the expectation value $\text{Exp}[1-I]$, and the shaded bands represent the standard deviations $\sqrt{\text{Var}[1-I]}$, $\sqrt{\text{Var}[1-I]/10}$, and $\sqrt{\text{Var}[1-I]/100}$.
The upper horizontal axis is the dimensionless variable $\Pi(\omega) d$, and the lower horizontal axis is $\omega_{\text{eV}} \equiv \omega / \text{eV}$ normalized by $n_{e0} \equiv n_e / \text{m}^{-3}$ and $d_{100\text{Mpc}} \equiv d / (100\,\text{Mpc})$, where $n_e$ is the plasma density and $d$ is the propagation distance. 
The right vertical axis is $1-I$ normalized by $P_{B*} k_*^2 d / (8\pi^2 M_{\rm P}^2)$. 
The left vertical axis is $1-I$ normalized by $B_{*\text{nG}}^2 \equiv (B_* / \text{nG})^2$, $d_{100\text{Mpc}}$, and $\lambda_{*\text{Mpc}} \equiv \lambda_{*} / \text{Mpc}$, where $B_{*}$ is the magnetic field strength at the peak scale $k_*$, and $\lambda_{*} \equiv 2\pi/k_*$ is the characteristic length.
In each row, the left and right panels show the same quantity: The vertical axis is linear in the left, but logarithmic in the right.
}
\label{fig:I}
\end{figure}

\begin{figure}
\centering
\includegraphics[width=0.45\linewidth]{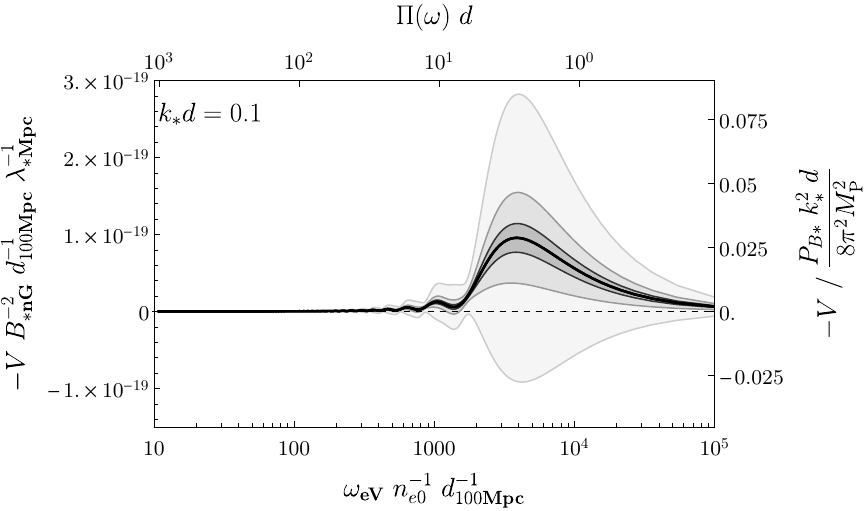}
\hskip 0.5cm
\includegraphics[width=0.45\linewidth]{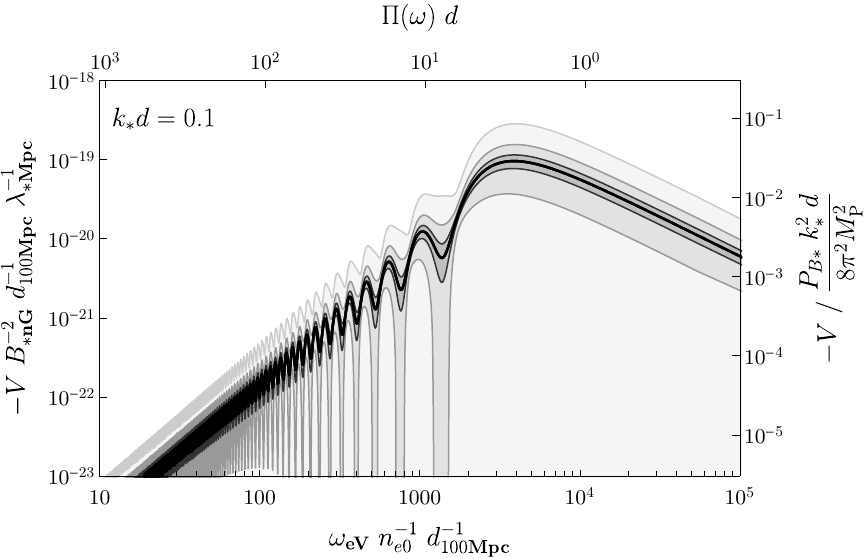}
\\
\includegraphics[width=0.45\linewidth]{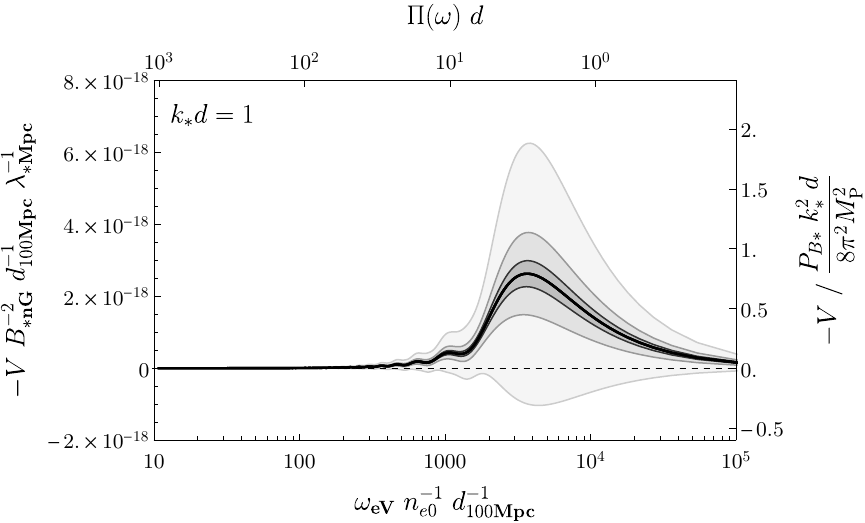}
\hskip 0.5cm
\includegraphics[width=0.45\linewidth]{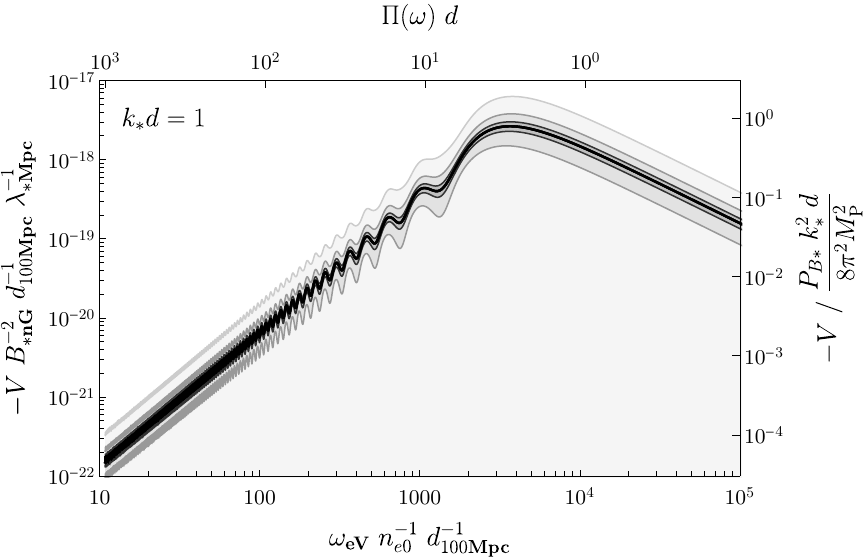}
\\
\includegraphics[width=0.45\linewidth]{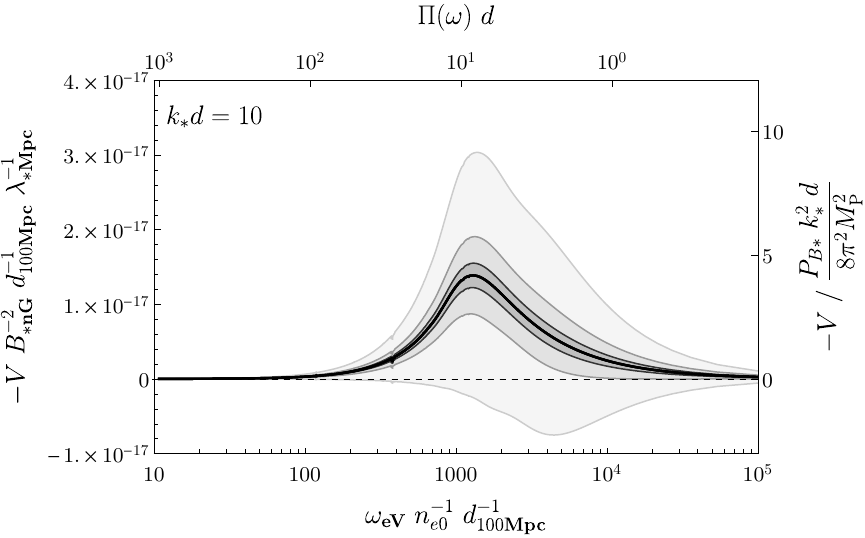}
\hskip 0.5cm
\includegraphics[width=0.45\linewidth]{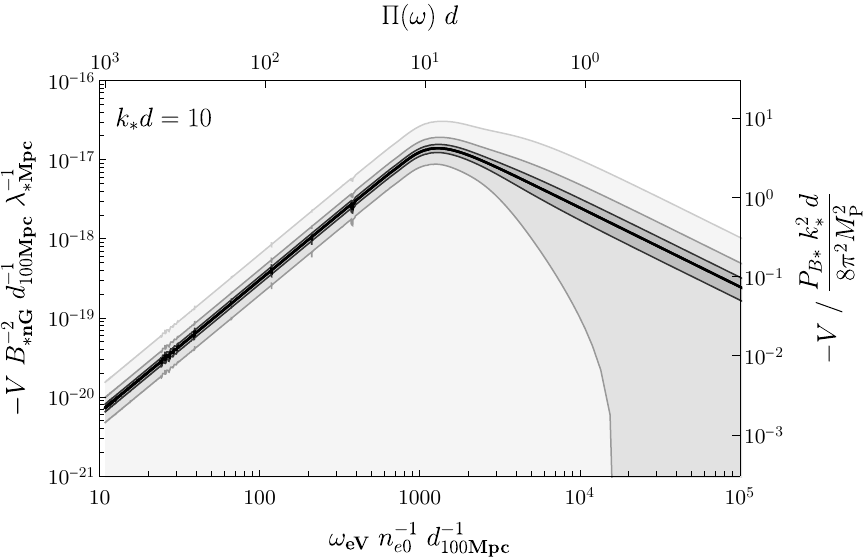}
\\
\includegraphics[width=0.45\linewidth]{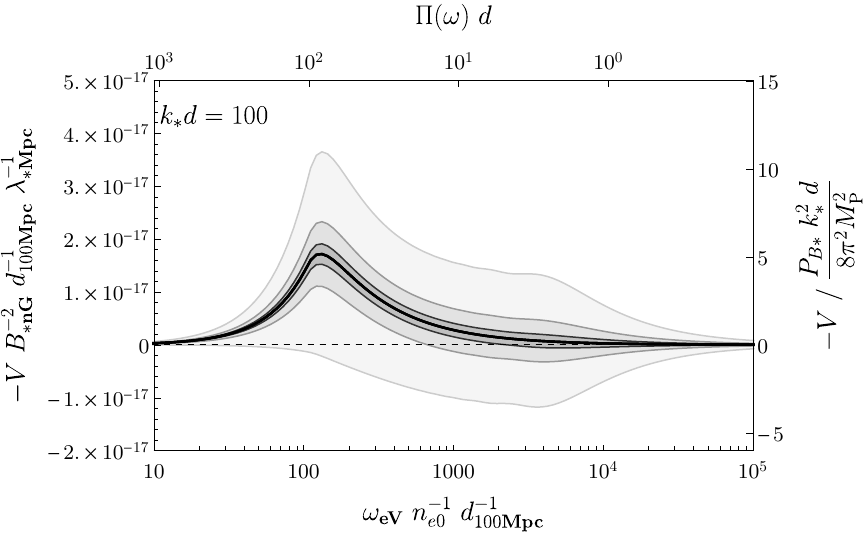}
\hskip 0.5cm
\includegraphics[width=0.45\linewidth]{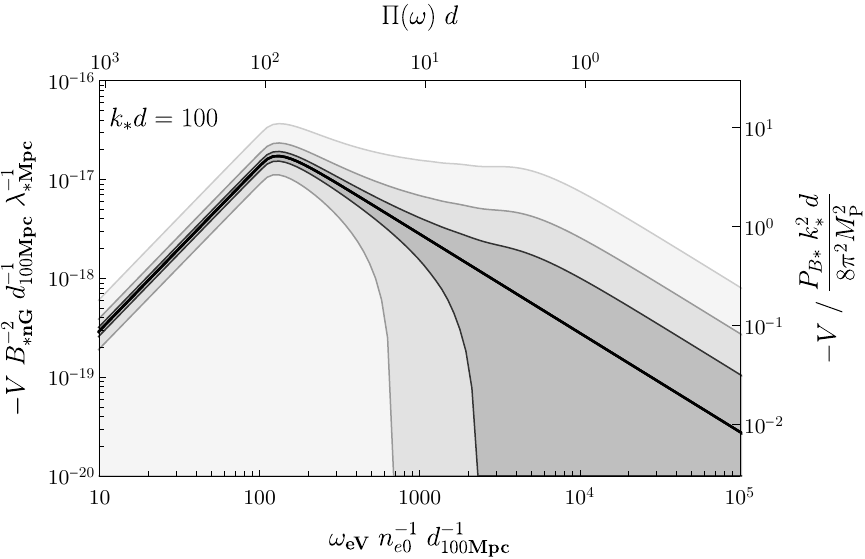}
\caption{\small
Stokes parameter $-V$ (circular polarization) of GWs with the maximally helical ($P_{aB*} = P_{B*}$) magnetic field power spectra \eqref{eq:PB} and \eqref{eq:PaB} is shown as a function of the angular frequency $\omega$ for various values of $k_* d$.
From top to bottom, the cases $k_* d = 0.1$, $1$, $10$, and $100$ are shown.
In each panel, the black line represents the expectation value $-\text{Exp}[V]$, and the shaded bands represent the standard deviations $\sqrt{\text{Var}[V]}$, $\sqrt{\text{Var}[V]/10}$, and $\sqrt{\text{Var}[V]/100}$.
The upper horizontal axis is the dimensionless variable $\Pi (\omega) d$, and the lower horizontal axis is $\omega_{\text{eV}} \equiv \omega / \text{eV}$ normalized by $n_{e0} \equiv n_e / \text{m}^{-3}$ and $d_{100\text{Mpc}} \equiv d / (100\,\text{Mpc})$, where $n_e$ is the plasma density and $d$ is the propagation distance. 
The right vertical axis is $-V$ normalized by $P_{B*} k_*^2 d / (8\pi^2 M_{\rm P}^2)$. 
The left vertical axis is $-V$ normalized by $B_{*\text{nG}}^2 \equiv (B_* / \text{nG})^2$, $d_{100\text{Mpc}}$, and $\lambda_{*\text{Mpc}} \equiv \lambda_{*} / \text{Mpc}$, where $B_{*}$ is the magnetic field strength at the peak scale $k_*$, and $\lambda_{*} \equiv 2\pi/k_*$ is the characteristic length.
In each row, the left and right panels show the same quantity: The vertical axis is linear in the left, but logarithmic in the right.
}
\label{fig:V}
\end{figure}

We finally present in Figs.~\ref{fig:I} and \ref{fig:V} the behavior of the Stokes parameters $I$ and $V$ as functions of $\Pi(\omega) d$ for various values of $k_* d$.
In these plots, the black lines represent the expectation values while the gray bands correspond to the standard deviations $\sqrt{\mathrm{Var}[1-I]}$ and $\sqrt{\mathrm{Var}[V]}$.
The three different bands are for $(1, 1/\sqrt{10}, 1/\sqrt{100})$ times the original one, respectively.
We plot these bands to indicate how the statistics improve as we increase the number of measurements.
In these gray bands, we envision a situation where we observe high-frequency GWs from different directions emitted from certain sources that can be regarded as almost identical.
Then the only difference in the observed signal is from the different realizations of the magnetic field, and the variance improves inversely to the number of observations.
Hence, each line can roughly be understood as $1$, $10$, and $100$ times of observations.

Regarding the behavior of the expectation values, we note that those of the intensity $I$ and circular polarization $V$ are solely determined from the integrals $\alpha$ and $\beta$, respectively, see Eq.~\eqref{eq:Stokes_statistical}.
Thus, the black lines in Figs.~\ref{fig:I} and \ref{fig:V} track the behavior of these integrals shown in Fig.~\ref{fig:abc}.
Indeed, the expectation value of the intensity switches from increasing with oscillations (small $\omega$) to constant (large $\omega$) at $\Pi d = \mathcal{O}(1)$ for $k_* d \ll 1$ (top row of Fig.~\ref{fig:I}), while it switches from increasing (small $\omega$) to constant (large $\omega$) at $\Pi d \approx k_* d$ for $k_* d \gg 1$ (bottom two rows of Fig.~\ref{fig:I}).
These behaviors are identical to those of $\alpha$ in the top panel of Fig.~\ref{fig:abc}.
Similarly, the expectation value of the circular polarization $V$ switches from increasing with oscillations (small $\omega$) to decreasing (large $\omega$) at $\Pi d = \mathcal{O}(1)$ for $k_* d \ll 1$ (top row of Fig.~\ref{fig:V}), while it switches from increasing (small $\omega$) to decreasing (large $\omega$) at $\Pi d \approx k_* d$ for $k_* d \gg 1$ (bottom two rows of Fig.~\ref{fig:V}).
Again, these behaviors are identical to those of $\beta$ in the middle panel of Fig.~\ref{fig:abc}.

In contrast, the behavior of the variances is somewhat nontrivial:
\begin{itemize}
\item 
The variance of $1 - I$ is determined by the combination $2 (\alpha^2 + \beta^2 + \gamma^2)$, see Eq.~\eqref{eq:Stokes_statistical}.
For $k_* d \lesssim 1$, only $\Pi d = \mathcal{O} (1)$ appears as a characteristic scale.
In the frequency range $\Pi d \gg 1$, the square root of the variance $\sqrt{\text{Var} [1 - I]}$ is of the same order as the expectation value $\text{Exp}[ 1 - I ]$, while in the range $\Pi d \ll 1$, the former approaches exactly the latter.
For $k_* d \gg 1$, on the other hand, a nontrivial structure appears in the frequency range $k_* d \gg \Pi d \gg 1$.
In this range, the variance is $2 (\alpha^2 + \beta^2 + \gamma^2) \approx 2 \alpha^2$, and hence the expectation value and the square root of the variance satisfies $\sqrt{\text{Var} [1 - I]} \approx \text{Exp} [1 - I] / \sqrt{2}$. 

\item 
The variance of $V$ is determined by the combination $2 (\alpha^2 + \beta^2 - \gamma^2)$, see Eq.~\eqref{eq:Stokes_statistical}.
For $k_* d \lesssim 1$, the square root of the variance $\sqrt{\text{Var} [V]}$ is comparable to the expectation value $\text{Exp} [V]$ for all the frequency range.
For $k_* d \gg 1$, on the other hand, these two are comparable only for $\Pi d \gg k_* d$.
In the opposite regime $\Pi d \ll k_* d$, the former dominates the latter.
Indeed, we confirm this behavior in Sec.~\ref{subsec:analytic}.
\end{itemize}

Finally, we comment on the relative sign between the parameter $V$ representing the circular polarization of GWs, and the parameter $P_{aB*}$ representing the helicity of the background magnetic field. 
In our convention, both $V$ and $P_{aB*}$ are defined such that a positive sign corresponds to right-handed helicity~\cite{Seto:2008sr, Durrer:2013pga}.
As is evident from Fig.~\ref{fig:V}, when the background magnetic field has a positive $P_{aB*}$, the GWs after conversion tend to acquire a negative $V$. 
Namely, the sign of $V$ generated through the conversion is opposite to that of $P_{aB*}$.
This can be intuitively understood as follows.
Assume that the background magnetic field has a positive $P_{aB*}$. 
From a microscopic perspective, the graviton-photon conversion is regarded as a process in which a single graviton interacts with a background photon to produce a single photon.
In the rest frame of the produced photon (which is accompanied by the plasma mass), the graviton and the background photon have momenta in opposite directions.
Although the background photon typically has positive (right-handed) helicity from assumption, it carries negative angular momentum with respect to the direction of propagation of the graviton, since its momentum is opposite to that of the graviton.
Then, a right-handed graviton carrying positive angular momentum (helicity $+2$) can interact with such a background photon to create a photon with helicity $+1$. On the other hand, a left-handed graviton cannot interact with the background photon to produce a photon from the conservation of angular momentum.
Given an initially unpolarized ensemble of gravitons (i.e., equal numbers of right- and left-handed modes), the above process selectively converts right-handed gravitons into right-handed photons, leaving left-handed gravitons unaffected. Hence, the resulting circular polarization of gravitons $V$ will be negative.

\subsection{Analytic interpretation}
\label{subsec:analytic}

The numerical results presented above indicate that the $\omega$-dependence of the Stokes parameters $I$ and $V$ exhibits distinct behaviors depending on the value of $k_* d$. 
Apparently, for $k_* d \ll 1$, the $\omega$-dependence of $I$ and $V$ changes around $\Pi d = \mathcal{O}(1)$ (top row of Figs.~\ref{fig:I} and \ref{fig:V}). 
In contrast, for $k_* d \gg 1$, the behavior of the expectation values of $I$ and $V$ appears to change around $\Pi d \approx k_* d$. 
Furthermore, looking at the variances, we see that $\Pi d = \mathcal{O}(1)$ also provides a characteristic threshold: In the regime $1 \lesssim \Pi d \lesssim k_* d$, both $\sqrt{\text{Var}[1-I]}$ and $\sqrt{\text{Var}[V]}$ remain approximately constant, whereas for $\Pi d \lesssim 1$, $\sqrt{\text{Var}[1-I]}$ increases and $\sqrt{\text{Var}[V]}$ decreases as $\Pi d$ becomes smaller (bottom two rows of Figs.~\ref{fig:I} and \ref{fig:V}).
Now we provide an analytic interpretation for these $\omega$-dependent behaviors.
To this end, we first analyze the structure of the kernels $\mathcal{C}_\alpha$, $\mathcal{C}_\beta$, and $\mathcal{C}_\gamma$ for $\Pi d \gg 1$ and $\Pi d \ll 1$ separately, and then investigate the behavior of $I$ and $V$.
Note that since $\Pi \propto \omega^{-1}$, the condition $\Pi d \gg 1$ corresponds to the low-frequency regime, while $\Pi d \ll 1$ corresponds to the high-frequency regime.

\subsubsection{\texorpdfstring{Low-frequency regime: $\Pi d \gg 1$}{Low-frequency regime: Delta d >> 1}}

In the low-frequency regime $\Pi d \gg 1$, the functions $(2 / \theta_{\pm}) \sin (\theta_{\pm} / 2)$ with $\theta_{\pm} = \Pi d \pm kd \cdot z$ contained in the kernels \eqref{eq:C} are replaced by 
\begin{align}
    \frac{2 \sin (\theta_{\pm} / 2)}{\theta_{\pm}}
    \approx 
    \begin{dcases}
        \frac{2}{\Pi d} \sin \frac{\Pi d}{2} \pm \frac{kd \cdot z}{\Pi d} \bigg( \cos \frac{\Pi d}{2} - \frac{2}{\Pi d} \sin \frac{\Pi d}{2} \bigg) \,,
        &(kd \ll 1) \\
        \frac{2}{\Pi d} \sin \bigg[ \frac{\Pi d}{2} \bigg( 1 \pm \frac{kd}{\Pi d} \,z \bigg) \bigg] \,,
        &(1 \lesssim kd \ll \Pi d) \\
        \frac{2\pi}{kd} \, \delta \bigg( z \pm \frac{\Pi d}{kd} \bigg) \,,
        &(kd \sim \Pi d)
    \end{dcases}
\end{align}
Here, for $kd \ll 1$, we performed the Taylor expansion with respect to the small parameter $kd$.
For $1 \lesssim kd \ll \Pi d$, we expanded with respect to $kd/\Pi d$.
For $kd \sim \Pi d$, we approximated the functions $(2 / \theta_{\pm}) \sin (\theta_{\pm} / 2)$ by the delta function, since they exhibit a sharp peak around $\theta_{\pm} = 0$.
Applying this replacement, the kernels $\mathcal{C}_\alpha$ and $\mathcal{C}_\beta$ for $\Pi d \gg 1$ are approximated as
\begin{align}
    \mathcal{C}_\alpha (\Pi d, kd)
    &\xrightarrow{\Pi d \gg 1}
    \begin{dcases}
        \frac{16}{3} \bigg( \frac{kd}{\Pi d} \bigg)^2 \sin^2 \bigg( \frac{\Pi d}{2} \bigg) \,, 
        &(kd \ll 1) \\
        4 \bigg( \frac{kd}{\Pi d} \bigg)^2 
        \bigg[ \frac{2}{3} - \frac{(kd)^2 \sin kd + kd \cos kd - \sin kd}{(kd)^3} \cos \Pi d \bigg] \,,
        &(1 \lesssim kd \ll \Pi d)\\
        \pi (kd) \bigg[ 1 + \bigg( \frac{\Pi d}{kd} \bigg)^2 \bigg] \,, 
        &(\Pi d < kd)
    \end{dcases}
    \label{Calpha_low}
    \\
    \mathcal{C}_\beta (\Pi d, kd)
    &\xrightarrow{\Pi d \gg 1}
    \begin{dcases}
        - \frac{4}{3} kd \bigg( \frac{kd}{\Pi d} \bigg)^2 \bigg( \sin \Pi d + \frac{2 \cos \Pi d - 2}{\Pi d} \bigg) \,, &(kd \ll 1) \\
        4  ( kd \cos kd - \sin kd ) \frac{\sin \Pi d}{(\Pi d)^2}\,,
        &(1 \lesssim kd \ll \Pi d) \\
        2\pi \, \Pi d \,. 
        &(\Pi d < kd)
    \end{dcases}
    \label{Cbeta_low}
\end{align}
These expressions reproduce the behavior of the kernels $\mathcal{C}_\alpha$ and $\mathcal{C}_\beta$ with $\Pi d =10$ and $100$ shown in Fig.~\ref{fig:C}.
In particular, these kernels are strongly suppressed when $kd \ll \Pi d$.
On the other hand, noting that the kernel $\mathcal{C}_\gamma$ contains the product of the oscillatory functions, $(2 / \theta_{+}) \sin (\theta_{+} / 2) \cdot (2 / \theta_{-}) \sin (\theta_{-} / 2)$, for $\Pi d \gg 1$ it is approximated as
\begin{align}
    \mathcal{C}_\gamma (\Pi d, kd) &\xrightarrow{\Pi d \gg 1} 
    \begin{dcases}
        \frac{16}{3} \bigg( \frac{kd}{\Pi d} \bigg)^2 \sin^2 \bigg( \frac{\Pi d}{2} \bigg)  \,,
        &(kd \ll 1) \\
        - 4 \bigg( \frac{kd}{\Pi d} \bigg)^2 \bigg[ \frac{2}{3} \cos \Pi d - \frac{(kd)^2 \sin kd + kd \cos kd - \sin kd}{(kd)^3} \bigg] \,,
        &(1 \lesssim kd \ll \Pi d)\\
        \pi (kd) \frac{\sin \Pi d}{\Pi d} \bigg[ 1 + \bigg( \frac{\Pi d}{kd} \bigg)^2 \bigg] \,.
        &(\Pi d < kd)
    \end{dcases}
    \label{Cgamma_low}
\end{align}
Here, for $\Pi d < kd$, we employed the approximation in which one of the oscillatory functions is replaced by the delta function.
Note that, when $\Pi d < kd$, $\mathcal{C}_\gamma$ is suppressed than $\mathcal{C}_\alpha$ by the large factor $\Pi d$, which is seen in Fig.~\ref{fig:C}.

Multiplying the above expressions for the kernels by the power spectrum $P_B(k)$ or $P_{aB}(k)$ and integrating them, we can obtain the analytic estimates for $\alpha$, $\beta$, and $\gamma$ in the low-frequency regime ($\Pi d \gg 1$).
We adopt the power-law forms for $P_B(k)$ and $P_{aB}(k)$ with a peak scale $k_*$ given by Eqs.~\eqref{eq:PB} and \eqref{eq:PaB} without using specific values for the spectral indices $n_l$, $n_h$, $n_{al}$, and $n_{ah}$ (but assuming $n_l > 0$,  $n_h < -3$, $n_{al} > 0$, and $n_{ah} < -3$ below).
For a given value of $\Pi d (\gg 1)$, we may distinguish cases based on whether $k_* d$ exceeds the threshold $kd \approx \Pi d$ at which the behavior of the kernels changes.
In the case of $k_* d \ll \Pi d$, the integral $\alpha$ is estimated as 
\begin{align}
    \alpha &\xrightarrow{\Pi d \gg 1}
        \, \sim \frac{P_{B*} k_*^2 d}{8\pi^2 M_{\rm P}^2} 
        \cdot \text{max}
        \bigg[
        \frac{k_* d}{(\Pi d)^2} , 
        \bigg( \frac{k_* d}{\Pi d} \bigg)^{-2 - {n_h}}
        \bigg]\,,
        \qquad (k_* d \ll \Pi d)
        \label{lowalpha1}
\end{align}
where some numerical factor is omitted.
On the other hand, when $\Pi d \ll k_* d$, we obtain 
\begin{align}
    \alpha &\xrightarrow{\Pi d \gg 1}
    \frac{P_{B*} k_*^2 d}{8\pi^2 M_{\rm P}^2}\,
    \pi \bigg( \frac{1}{{n_l}+2} - \frac{1}{{n_h}+2} \bigg) \,.
    \qquad (1 \ll \Pi d \ll k_* d)
    \label{lowalpha3}
\end{align}
Similarly, for $k_* d \ll \Pi d$, the integral $\beta$ is estimated as
\begin{align}
    \beta \xrightarrow{\Pi d \gg 1}
    \, \sim 
    \frac{P_{aB*} k_*^2 d}{8\pi^2 M_{\rm P}^2} \times
    \begin{dcases}
        \bigg( \frac{k_*d}{\Pi d} \bigg)^{\text{min}(2, -2-n_{ah})} \,,
        &(k_* d \ll 1 \ll \Pi d) \\
        \text{max} \bigg[ \frac{1}{(\Pi d)^2} , \bigg(\frac{k_* d}{\Pi d}\bigg)^{-2-n_{ah}} \bigg] \,,
        &(1 \ll k_* d \ll \Pi d)
    \end{dcases}
    \label{lowbeta1}
\end{align}
and, for $\Pi d \ll k_* d$, we obtain
\begin{align}
    \beta \xrightarrow{\Pi d \gg 1}
    \frac{P_{aB*} k_*^2 d}{8\pi^2 M_{\rm P}^2}\,
    2\pi 
        \bigg( \frac{1}{{n_{al}} + 1} - \frac{1}{{n_{ah}} + 1} \bigg) \frac{\Pi d}{k_* d} \,.
        &\qquad (1\ll \Pi d \ll k_* d)
        \label{lowbeta3}
\end{align}
Finally, the integral $\gamma$ is estimated as 
\begin{align}
    \gamma \xrightarrow{\Pi d \gg 1}
    \frac{P_{B*} k_*^2 d}{8\pi^2 M_{\rm P}^2} \times 
    \begin{dcases}
        \sim \frac{k_* d}{(\Pi d)^2} \,,
        &(k_* d \ll \Pi d) \\
        \pi \bigg( \frac{1}{{n_l}+2} - \frac{1}{{n_h}+2} \bigg) \frac{\sin \Pi d}{\Pi d} \,.
        & (1\ll \Pi d \ll k_* d)
    \end{dcases}
    \label{lowgamma3}
\end{align}
Setting the spectral indices as $n_l = 2$, $n_{al} = 3$ and $n_{h} = n_{ah} =  -11/3$, and focusing on the case of $k_* d \gg 1$, the leading behaviors of $\alpha$ and $\beta$ are estimated as $\alpha \propto \beta \propto \omega^{5/3}$ for $k_* d \ll \Pi d$, while $\alpha \propto \omega^{0}$ and $\beta \propto \omega^{-1}$ for $1 \ll \Pi d \ll k_* d$. 
These analytic estimates well reproduce the behaviors of $\alpha$ and $\beta$ in the low-frequency regime ($\Pi d \gg 1$) shown in the top and middle rows of Fig.~\ref{fig:abc}.

Based on the above expressions for $\alpha$, $\beta$, and $\gamma$, let us analyze the behavior of the Stokes parameters $I$ and $V$ in the low-frequency regime.  
We discuss the cases $k_* d \ll 1$ and $k_* d \gg 1$ separately.
First, when $k_* d \ll 1$, it always follows that $k_* d \ll \Pi d$ in the low-frequency regime ($\Pi d \gg 1$). Hence, we obtain the following analytic formulas for the expectation values in the low-frequency limit (supposing $|n_{h}|, |n_{ah}| < 4$),
\begin{align}
    k_* d \ll 1: ~
    \text{Exp}[1-I] 
    = 2 \alpha 
    &\sim \frac{P_{B*} k_*^2 d}{8 \pi^2 M_{\rm P}^2} 
     \bigg( \frac{k_*d}{\Pi d} \bigg)^{-2-n_{h}}
    \propto \omega^{-2-n_{h}}\,,
    \qquad (\Pi d \to \infty) \\
    - \text{Exp}[V] 
    = 2 \beta 
    &\sim \frac{P_{aB*} k_*^2 d}{8\pi^2 M_{\rm P}^2} \bigg( \frac{k_*d}{\Pi d} \bigg)^{-2-n_{ah}}
    \propto \omega^{-2-n_{ah}} \,,
    \qquad (\Pi d \to \infty) 
\end{align}
Although all oscillatory factors are omitted in these formulas, their overall $\omega$-dependence in the low-frequency regime is consistent with the numerical results presented in the top panel of Figs.~\ref{fig:I} and \ref{fig:V} ($k_* d = 0.1$).

In contrast, when $k_* d \gg 1$, we obtain the analytic formulas (with $|n_{h}|, |n_{ah}| < 4$) as
\begin{align}
    k_* d \gg 1:~
    \text{Exp}[1-I] 
    &\approx \frac{P_{B*} k_*^2 d}{8 \pi^2 M_{\rm P}^2} \times 
    \begin{dcases}
        \sim 
        \bigg( \frac{k_* d}{\Pi d} \bigg)^{-2 - {n_h}}
        \propto \omega^{- 2 - {n_h}}
        \,, 
        &(k_* d \ll \Pi d)        
        \\
        2 \pi
        \bigg( \frac{1}{{n_l}+2} - \frac{1}{{n_h}+2}
        \bigg)
        \propto \omega^0
        \,,
        &(1 \ll \Pi d \ll k_* d)   
    \end{dcases}
    \label{low_ExpI_largeks}
    \\
    -\text{Exp}[V] 
    &\approx \frac{P_{aB*} k_*^2 d}{8 \pi^2 M_{\rm P}^2} \times
    \begin{dcases}
        \sim \bigg( \frac{k_*d}{\Pi d} \bigg)^{-2-n_{ah}}
        \propto \omega^{-2-n_{ah}}
        \,,
        &(k_* d \ll \Pi d) \\
        4\pi 
        \bigg( \frac{1}{{n_{al}} + 1} - \frac{1}{{n_{ah}} + 1} \bigg) \frac{\Pi d}{k_* d} 
        \propto \omega^{-1}
        \,.
        &(1 \ll \Pi d \ll k_* d)
    \end{dcases}
    \label{low_ExpV_largeks}
\end{align}
These formulas reproduce the results in the low-frequency regime ($\Pi d \gg 1$) shown in the bottom two panels of Figs.~\ref{fig:I} and \ref{fig:V} ($k_* d = 10$ and $100$), where $n_l = 2$, $n_{al} = 3$, and $n_{h} = n_{ah} = -11/3$.
There, we see that $\text{Exp}[1-I] \propto \omega^{-2-n_h} = \omega^{5/3}$ and $-\text{Exp}[V] \propto \omega^{-2-n_{ah}} = \omega^{5/3}$ for $k_* d \ll \Pi d$, whereas  $\text{Exp}[1-I] \approx \text{const.}$ and $-\text{Exp}[V] \propto \omega^{-1}$ for $1 \ll \Pi d \ll k_* d$. 
Note that $-\text{Exp}[V]$ is peaked at $\Pi d \approx k_* d$, and the peak value is predicted as $-\text{Exp}[V] \sim P_{aB*} k_*^2 d / (8 \pi^2 M_{\rm P}^2)$.
In fact, that peak position can be expected from the behavior of the kernel $\mathcal{C}_\beta$ shown in Fig.~\ref{fig:C}: We observe that $\mathcal{C}_\beta$ is significantly suppressed when $kd \lesssim \Pi d$ for $\Pi d \gg 1$. Thus, if $k_*d \gg 1$, the convolution can take a large value only in the frequency range $\Pi d \lesssim k_* d$.
Additionally, since the asymptotic value of $\mathcal{C}_\beta$ in large $kd$ scales proportionally to $\Pi d$, the convolution takes the maximum value at the frequency $\Pi d \approx k_* d$.

Finally, let us briefly discuss the variance of the Stokes parameters $I$ and $V$ in the low-frequency regime $(\Pi d \gg 1)$.
As shown in Eq.~\eqref{eq:Stokes_statistical}, the variances are generally given by $\text{Var}[1-I] = 2(\alpha^2 + \beta^2 + \gamma^2)$ and $\text{Var}[V] = 2(\alpha^2 + \beta^2 - \gamma^2)$.
From Eqs.~\eqref{lowalpha1}, \eqref{lowbeta1}, and \eqref{lowgamma3}, we see that $\gamma$ is more suppressed than $\alpha$ and $\beta$ in the low-frequency limit, $\Pi d \to \infty$, if $|n_h|, |n_{ah}| < 4$.
Thus, we can estimate the variances as $\sqrt{\text{Var}[1-I]} \approx \sqrt{\text{Var}[V]} \approx \sqrt{2(\alpha^2 + \beta^2)}$ in this limit.
Moreover, if the maximally helical case ($|P_{aB*}| = P_{B*}$) is considered, we obtain $\alpha \sim \beta$ and thus $\sqrt{\text{Var}[1-I]} \approx \sqrt{\text{Var}[V]} \sim 2 \alpha \approx \text{Exp}[1-I]$ in the limit $\Pi d \to \infty$, which is consistent with the results shown in Figs.~\ref{fig:I} and \ref{fig:V}.
Meanwhile, in the case of $k_* d \gg 1$, the variances show a distinct behavior within a particular frequency range $1 \ll \Pi d \ll k_* d$.
In the bottom two panels of Figs.~\ref{fig:I} and \ref{fig:V} ($k_* d = 10$ and $100$), we observe that the variance of $I$ and $V$ remains almost constant within that frequency range. This can be interpreted analytically as follows.
Comparing Eqs.~\eqref{lowalpha3}, \eqref{lowbeta3}, and \eqref{lowgamma3}, we see that $\alpha$ is much larger than $\beta$ and $\gamma$ in that range $1 \ll \Pi d \ll k_* d$. Hence, the formula for the variances \eqref{eq:Stokes_statistical} can be approximated as 
\begin{align}
    k_* d \gg 1: ~ 
    \sqrt{\text{Var}[1-I]} \approx \sqrt{\text{Var}[V]} \approx \sqrt{2} \,\alpha \approx \frac{1}{\sqrt{2}} \text{Exp}[1-I]\,,
    \qquad (1 \ll \Pi d \ll k_* d)
\end{align}
which is almost constant in that range from Eq.~\eqref{lowalpha3}.

\subsubsection{\texorpdfstring{High-frequency regime: $\Pi d \ll 1$}{High-frequency regime: Delta d << 1}}

In the high-frequency regime $\Pi d \ll 1$, we perform Taylor expansion with respect to $\Pi d$.
We write the expansion of the kernels $\mathcal{C}_i$ $(i = \alpha, \beta, \gamma)$ as 
\begin{align}
    \mathcal{C}_\alpha (\Pi d, kd) 
    &= \mathcal{C}_\alpha^{(0)}(kd) + \mathcal{C}_\alpha^{(2)}(kd) \cdot (\Pi d)^2 + \cdots \,, 
    \label{expandCa} \\
    \mathcal{C}_\beta (\Pi d, kd) 
    &= \mathcal{C}_\beta^{(1)}(kd) \cdot \Pi d + \cdots \,, 
    \label{expandCb} \\
    \mathcal{C}_\gamma (\Pi d, kd) 
    &= \mathcal{C}_\gamma^{(0)}(kd) + \mathcal{C}_\gamma^{(2)}(kd) \cdot (\Pi d)^2 + \cdots \,.
    \label{expandCc}
\end{align}
Note that $\mathcal{C}_\alpha$ and $\mathcal{C}_\gamma$ are even while $\mathcal{C}_\beta$ is odd with respect to $\Pi d$, as can be seen from their definition \eqref{eq:C}.
We should also note that $\mathcal{C}_\alpha^{(0)}(kd) = \mathcal{C}_\alpha(\Pi d=0, kd)$ and $\mathcal{C}_\gamma^{(0)}(kd) = \mathcal{C}_\gamma(\Pi d=0, kd)$ are identical since we have $(2/\theta_-) \sin (\theta_- / 2) = (2/\theta_+) \sin (\theta_+ / 2) $ when $\Pi d = 0$.
The expansion coefficients are obtained as
\begin{align}
    \mathcal{C}_{\alpha}^{(0)}(kd) = \mathcal{C}_{\gamma}^{(0)}(kd) 
    &= 2 \bigg( \cos kd - \frac{\sin kd}{kd} + kd \, \text{Si} (kd) \bigg) \,, \\
    \mathcal{C}_{\alpha}^{(2)}(kd) 
    &= 2 \bigg( \frac{4 \cos kd - 4}{(kd)^2} + \frac{\sin kd + \text{Si}(kd)}{kd} \bigg) \,, \\
    \mathcal{C}_{\beta}^{(1)}(kd)
    &= 4 \bigg( \frac{2 \cos kd -2}{kd} + \text{Si}(kd) \bigg) \,, \\
    \mathcal{C}_{\gamma}^{(2)}(kd)
    &= \frac{1}{3} \bigg(
    \frac{8 \cos kd - 8}{(kd)^2} - \frac{\sin kd - 6 \, \text{Si}(kd)}{kd} - \cos kd - kd \, \text{Si}(kd) \bigg)
    \,,
\end{align}
where $\text{Si}(x) \equiv \int_0^x dt \, (\sin t)/t$.
The asymptotic behaviors of the leading coefficients $\mathcal{C}^{(0)}_\alpha = \mathcal{C}^{(0)}_\gamma$ and $\mathcal{C}^{(1)}_\beta$ in $kd \ll 1$ and $kd \gg 1$ are given by 
\begin{align}
    \mathcal{C}^{(0)}_\alpha(kd) = \mathcal{C}^{(0)}_\gamma(kd) 
    &\approx 
    \begin{dcases}
        \frac{4}{3}(kd)^2 \,, &(kd \ll 1) \\
        \pi (kd) \,, &(kd \gg 1) 
    \end{dcases} 
    \\
    \mathcal{C}^{(1)}_\beta(kd) 
    &\approx 
    \begin{dcases}
        \frac{1}{9} (kd)^3 \,, &(kd \ll 1) \\
        2\pi \,. &(kd \gg 1) 
    \end{dcases} 
\end{align}
These asymptotic behaviors are manifested in the black curves ($\Pi d = 0.1$) shown in Fig.~\ref{fig:C}.

Similarly, we consider the Taylor expansion for the integrals $\alpha$, $\beta$, and $\gamma$:
\begin{align}
    \alpha &= \alpha_{(0)} + \alpha_{(2)} \cdot (\Pi d)^2 + \cdots \,, 
    \label{expand_a} \\
    \beta &= \beta_{(1)} \cdot \Pi d + \cdots \,,
    \label{expand_b} \\
    \gamma &= \gamma_{(0)} + \gamma_{(2)} \cdot (\Pi d)^2 + \cdots \,.
    \label{expand_c}
\end{align}
From the definition of $\alpha$, $\beta$, and $\gamma$ in Eq.~\eqref{eq:abcd}, the expansion coefficients are given by 
\begin{align}
    \alpha_{(0)} = \frac{1}{8\pi^2 M_{\rm P}^2} \int_0^\infty dk \, P_{B}(k) \, \mathcal{C}_\alpha^{(0)} (kd) \,, \qquad
    \beta_{(1)} = \frac{1}{8\pi^2 M_{\rm P}^2} \int_0^\infty dk \, P_{aB}(k) \, \mathcal{C}_\beta^{(1)} (kd) \,,
\end{align}
and so on. Note that the equality $\alpha_{(0)} = \gamma_{(0)}$ holds from $\mathcal{C}^{(0)}_\alpha = \mathcal{C}^{(0)}_\gamma$.

Equations \eqref{expand_a}--\eqref{expand_c} indicate that, in the high-frequency limit ($\Pi d \ll 1$), both $\alpha$ and $\gamma$ asymptotically approach the same constant value $\alpha_{(0)} =\gamma_{(0)}$, whereas $\beta$ decays as $\beta_{(1)} \cdot \Pi d \propto \omega^{-1}$. 
These behaviors in the high-frequency regime can be observed in Fig.~\ref{fig:abc}.
The values of $\alpha_{(0)} (=\gamma_{(0)})$ and $\beta_{(1)}$ depend on the shape of the power spectra $P_B(k)$ and $P_{aB}(k)$.
For $P_B(k)$ and $P_{aB}(k)$, below we assume the single broken power-law forms with a peak scale $k_*$ given by Eqs.~\eqref{eq:PB} and \eqref{eq:PaB}, without using specific values for the spectral indices $n_l$, $n_h$, $n_{al}$, and $n_{ah}$ (but assuming $n_l > 0$,  $n_h < -3$, $n_{al} > 0$, and $n_{ah} < -3$).
Let us separately discuss the two limiting cases, $k_* d \ll 1$ and $k_* d \gg 1$.

In the case of $k_* d \ll 1$, we consider Taylor expansion with respect to $k_* d$. We obtain
\begin{align}
    k_* d \ll 1: ~
    \alpha_{(0)} = \gamma_{(0)} &\approx \frac{P_{B*} k_*^2 d}{8\pi^2 M_{\rm P}^2} \frac{4}{3} \bigg( \frac{1}{n_l + 3} - \frac{1}{n_h + 3} \bigg) k_* d \,,
    \label{a0_smallks}
    \\
    \beta_{(1)} &\approx \frac{P_{aB*} k_*^2 d}{8\pi^2 M_{\rm P}^2} 
    \cdot \text{max} \bigg[ 
    \frac{1}{9} \bigg( \frac{1}{n_{al} + 4} - \frac{1}{n_{ah} + 4} \bigg) (k_* d)^2 , 
    \frac{4 (n_{ah} + 2)}{n_{ah} + 1} \Gamma(n_{ah}) \cos \bigg( \frac{\pi n_{ah}}{2} \bigg) (k_* d)^{-n_{ah} - 2}
    \bigg] \,,
    \label{b1_smallks}
\end{align}
which give the leading terms when $n_h < -3$.
Additionally, for $k_* d \ll 1$, the coefficients $\alpha_{(2)}$ and $\gamma_{(2)}$ are expanded as 
\begin{align}
    k_* d \ll 1:~
    \alpha_{(2)} &\approx \frac{P_{B*} k_*^2 d}{8\pi^2 M_{\rm P}^2} 
    \bigg[ - \frac{1}{9}\bigg( \frac{1}{n_l+3} - \frac{1}{n_h+3} \bigg) k_* d \bigg] \,, 
    \label{a2_smallks}\\
    \gamma_{(2)} &\approx \frac{P_{B*} k_*^2 d}{8\pi^2 M_{\rm P}^2} 
    \bigg[ - \frac{1}{9}\bigg( \frac{1}{n_l+3} - \frac{1}{n_h+3} \bigg) k_* d \bigg] \,,
    \label{c2_smallks}
\end{align}
which are, again, the leading terms when $n_h < -3$.
These will be used in the analysis of the variance of the Stokes parameters.

On the other hand, for $k_* d \gg 1$, we obtain the asymptotic forms as 
\begin{align}
    k_* d \gg 1:~
    \alpha_{(0)} = \gamma_{(0)} &\approx 
        \frac{P_{B*} k_*^2 d}{8\pi^2 M_{\rm P}^2}\, \pi \bigg( \frac{1}{n_l +2} - \frac{1}{n_h +2} \bigg) \,,
        \label{a0_largeks}\\
    \beta_{(1)} &\approx 
         \frac{P_{aB*} k_*^2 d}{8\pi^2 M_{\rm P}^2} \, 2\pi \bigg( \frac{1}{n_{al} + 1} - \frac{1}{n_{ah} + 1} \bigg) \frac{1}{k_* d} \,,
         \label{b1_largeks}\\
    \alpha_{(2)} &\approx \frac{P_{B*} k_*^2 d}{8\pi^2 M_{\rm P}^2} \, \pi \bigg( \frac{1}{n_l} - \frac{1}{n_h} \bigg) \frac{1}{(k_*d)^2} \,, 
    \label{a2_largeks}\\
    \gamma_{(2)} &\approx - \frac{P_{B*} k_*^2 d}{8\pi^2 M_{\rm P}^2} \frac{\pi}{6} \bigg( \frac{1}{n_l + 2} - \frac{1}{n_h + 2} \bigg) \,.
    \label{c2_largeks}
\end{align}

In terms of $\alpha$, $\beta$, and $\gamma$ investigated above, we can study the behavior of the Stokes parameters $I$ and $V$ in the high-frequency regime $(\Pi d \ll 1)$.
First, the expectation value of intensity $I$ is obtained as
\begin{align}
    \text{Exp}[1-I] = 2\alpha \approx 2 \alpha_{(0)} + 2 \alpha_{(2)} \cdot (\Pi d)^2 \,, 
    \qquad (\Pi d \ll 1)
    \label{high_expI}
\end{align}
where the constant $\alpha_{(0)}$ is given by Eq.~\eqref{a0_smallks} for $k_* d \ll 1$ and by Eq.~\eqref{a0_largeks} for $k_* d \gg 1$.
In the high-frequency limit ($\Pi d \to 0$), $\text{Exp}[1-I]$ converges to a constant value $2\alpha_{(0)}$.
In contrast, the expectation value of circular polarization $V$ decays in proportion to $\omega^{-1}$ in the high-frequency regime:
\begin{align}
    - \text{Exp}[V] = 2\beta \approx 2 \beta_{(1)} \cdot \Pi d \propto \omega^{-1}\,, 
    \qquad (\Pi d \ll 1)
    \label{high_expV}
\end{align}
where the coefficient $\beta_{(1)}$ is given by Eq.~\eqref{b1_smallks} for $k_* d \ll 1$ and by Eq.~\eqref{b1_largeks} for $k_* d \gg 1$.
Note that, in the case of $k_* d \gg 1$, the leading behaviors of $\text{Exp}[1-I]$ and $-\text{Exp}[V]$ in the high-frequency regime $\Pi d \ll 1$ agree with those in the lower frequency regime $1 \ll \Pi d \ll k_* d$ obtained in Eqs.~\eqref{low_ExpI_largeks} and \eqref{low_ExpV_largeks}.

We can further proceed to analyze the variances of the Stokes parameters in the high-frequency regime $(\Pi d \ll 1)$.
From Eq.~\eqref{eq:Stokes_statistical}, the variances of $I$ and $V$ are given by $\text{Var}[1-I] = 2(\alpha^2 + \beta^2 + \gamma^2)$ and $\text{Var}[V] = 2(\alpha^2 + \beta^2 - \gamma^2)$. 
Substituting the Taylor expansion \eqref{expand_a}--\eqref{expand_c}, we obtain 
\begin{align}
    \text{Var}[1-I] &= (2\alpha_{(0)})^2 + 2 ( 2 \alpha_{(0)} \alpha_{(2)} + 2 \alpha_{(0)} \gamma_{(2)} + \beta_{(1)}^2 ) (\Pi d)^2 + \cdots \,,
    \label{high_varI}\\
    \text{Var}[V] &=  [ 4 \alpha_{(0)} (\alpha_{(2)} - \gamma_{(2)}) + 2\beta_{(1)}^2 ] (\Pi d)^2 + \cdots \,,
    \label{high_varV}
\end{align}
where we used $\alpha_{(0)} = \gamma_{(0)}$.
Consequently, the standard deviations are given by 
\begin{align}
    \sqrt{\text{Var}[1-I]} &= 2\alpha_{(0)} + \bigg( \alpha_{(2)} + \gamma_{(2)} + \frac{\beta_{(1)}^2}{2\alpha_{(0)}} \bigg) (\Pi d)^2 + \cdots \,,
    \label{high_devI}\\
    \sqrt{\text{Var}[V]} &= \sqrt{4 \alpha_{(0)} (\alpha_{(2)} - \gamma_{(2)}) + 2 \beta_{(1)}^2 } \cdot (\Pi d) + \cdots \,.
    \label{high_devV}
\end{align}

For the intensity $I$, Eqs. \eqref{high_expI} and \eqref{high_devI} indicate that $\text{Exp}[1-I]$ and $\sqrt{\text{Var}[1-I]}$ converge to the same value $2\alpha_{(0)}$ in the high-frequency limit $\Pi d \to 0$.
Additionally, the coefficient of the $(\Pi d)^2$ term in Eq.~\eqref{high_devI} is approximately given by 
\begin{align}
     \alpha_{(2)} + \gamma_{(2)} + \frac{\beta_{(1)}^2}{2\alpha_{(0)}}
     \approx 
     \begin{dcases}
        \alpha_{(2)} + \gamma_{(2)} \approx 
        - \frac{P_{B*} k_*^2 d}{8\pi^2 M_{\rm P}^2} \frac{2}{9}\bigg( \frac{1}{n_l+3} - \frac{1}{n_h+3} \bigg) k_* d
        \, (\approx 2\alpha_{(2)})\,, 
        &(k_* d \ll 1) \\
        \gamma_{(2)} \approx 
        - \frac{P_{B*} k_*^2 d}{8\pi^2 M_{\rm P}^2} \frac{\pi}{6} \bigg( \frac{1}{n_l + 2} - \frac{1}{n_h + 2} \bigg)\,.
        &(k_* d \gg 1)    
     \end{dcases}
     \label{high_devI2}
\end{align}
Here, Eqs.~\eqref{a0_smallks}--\eqref{c2_smallks} are used for $k_* d \ll 1$ (with assumptions $n_h < -3$ and $n_{ah} < -3$), and Eqs.~\eqref{a0_largeks}--\eqref{c2_largeks} are used for $k_* d \gg 1$. 
When the coefficient \eqref{high_devI2} is negative (as in the case with $n_l = 2$, $n_{al} = 3$, and $n_h = n_{ah} = -11/3$), it follows that $\sqrt{\text{Var}[1-I]}$ decreases towards the low-frequency side. 
Indeed, the gray bands in the regime $\Pi d \ll 1$ of Fig.~\ref{fig:I} exhibit such a behavior.
Note that, for $k_* d \ll 1$, the relation $\sqrt{\text{Var}[1-I]} \approx \text{Exp}[1-I]$ holds since we have $\sqrt{\text{Var}[1-I]} = 2\alpha_{(0)} + 2\alpha_{(2)} (\Pi d)^2$ up to the order shown in Eq.~\eqref{high_devI2}.

For the circular polarization $V$, Eqs.~\eqref{high_expV} and \eqref{high_devV} indicate that both $-\text{Exp}[V]$ and $\sqrt{\text{Var}[V]}$ behave as $\propto \omega^{-1}$ in the high-frequency regime. 
In particular, in the case of $k_* d \gg 1$, the coefficient in Eq.~\eqref{high_devV} is approximately given by 
\begin{align}
    \sqrt{4 \alpha_{(0)} (\alpha_{(2)} - \gamma_{(2)}) + 2 \beta_{(1)}^2 }
    \approx 2\sqrt{|\alpha_{(0)} \gamma_{(2)}|}
    \approx \frac{P_{B*} k_*^2 d}{8\pi^2 M_{\rm P}^2} \frac{2 \pi}{\sqrt{6}} \bigg( \frac{1}{n_l + 2} - \frac{1}{n_h + 2} \bigg)\,, 
    \qquad
    (k_* d \gg 1)
    \label{high_devV1}
\end{align}
which holds for $n_h < -3$ and $n_{ah} < -3$.
We see that, in the case of $k_* d \gg 1$, $\sqrt{\text{Var}[V]}$ becomes larger than $-\text{Exp}[V] \approx 2\beta_{(1)} \Pi d$ by the factor $k_*d$ (compare Eqs.~\eqref{b1_largeks} and \eqref{high_devV1}). 
This provides an analytic interpretation for the trend observed in Fig.~\ref{fig:V}, where the width of the gray band in the high-frequency regime $(\Pi d \ll 1)$ increases as $k_* d$ increases.

\section{Discussion and Conclusions}
\label{sec:conclusions}

In this paper, we study the conversion between GWs and electromagnetic waves in the presence of stochastic magnetic fields, having cosmic magnetic fields in mind.
When GWs propagate through a magnetic field, they are converted into electromagnetic waves due to the universal coupling predicted by general relativity.
If the magnetic field is stochastically realized, a statistical framework is required to describe the conversion, which we work on in the present paper.
Following this approach, we compute the expectation values and variances of observables associated with GWs affected by the conversion, using ensemble averages over Gaussian magnetic field realizations.
Furthermore, we take into account the possible presence of magnetic field helicity, which is expected to affect the circular polarization of GWs.
By searching for these imprints encoded in GWs through observations, we may probe the properties of cosmic magnetic fields and gain insights into the underlying physics of their origin.

In Sec.~\ref{sec:review}, we derive the propagation equations for the graviton-photon system and solve them to first order in the magnetic field (Born approximation).
To describe the polarization state of GWs, we introduce the associated Stokes parameters and investigate how they are affected by the magnetic field in which they propagate.
In Sec.~\ref{sec:random}, we derive expressions for the expectation values and variances of the GW Stokes parameters in terms of the magnetic field power spectrum, assuming a Gaussian magnetic field and an initial state given by unpolarized GWs.
As shown in Eq.~\eqref{eq:Stokes_statistical}, these quantities are compactly expressed in terms of four basic components, $\alpha$, $\beta$, $\gamma$, and $\delta$, where $\delta$ vanishes identically.
Each component is expressed as a convolution of the magnetic field power spectrum with its own kernel (see Eq.~\eqref{eq:abcd}).
We find that nonzero signals appear in the intensity $I$ and circular polarization $V$, but not in linear polarizations $Q$ and $U$, when the initial state of GWs is unpolarized.
We also obtain a consistency relation \eqref{eq:consistency} between the GW intensity $I$ and circular polarization $V$.

In Sec.~\ref{sec:numerical}, using the magnetic field power spectra given by Eqs.~\eqref{eq:PB} and \eqref{eq:PaB} as an example, we examine the behavior of the GW Stokes parameters as functions of the parameters in this system.
The parameters are the effective mass difference between gravitons and photons $\Pi = \Pi_\gamma - \Pi_g = m_{\rm pl}^2 / (2 \omega)$ (with the photon plasma mass $m_{\rm pl}$ and the angular frequency of the waves $\omega$), the propagation distance $d$, and the typical wavenumber of the magnetic field $k_*$.
Among these, two dimensionless combinations $\Pi d$ and $k_* d$ can be constructed, which characterize the observables.
In Sec.~\ref{subsec:C}, we first show the behavior of the convolution kernels as functions of $\Pi d$. 
In Sec.~\ref{subsec:abc}, we numerically investigate the dependence on $\Pi d$ (or the angular frequency $\omega$) of the integrals $\alpha$, $\beta$, and $\gamma$ for various values of $k_* d$.
Using these results, in Sec.~\ref{subsec:NumCalcStokes}, we show the behavior of the expectation values and variances of the GW Stokes parameters $I$ and $V$.
We find that the expectation value of the conversion probability $\text{Exp}[1-I]$ changes its behavior from increasing with oscillation (small $\omega$) to constant (large $\omega$) at $\Pi d = \mathcal{O}(1)$ for $k_* d \ll 1$, whereas from increasing (small $\omega$) to constant (large $\omega$) at $\Pi d \approx k_* d$ for $k_* d \gg 1$ (see Fig.~\ref{fig:I}).
On the other hand, the variance $\text{Var}[1-I]$ shows a more curious $\omega$-dependence.
Specifically, we find that the particular frequency range $1 \lesssim \Pi d \lesssim k_* d$ has a variance relatively smaller than the expectation value (see the bottom panels of Fig.~\ref{fig:I}).
Thanks to the narrower variance, this frequency range can be favorable for observing signatures of the conversion.
For the circular polarization $V$, we find that the expectation value $-\text{Exp}[V]$ switches from increasing with oscillation (small $\omega$) to decreasing (large $\omega$) at $\Pi d = \mathcal{O}(1)$ for $k_* d \ll 1$, while from increasing (small $\omega$) to decreasing (large $\omega$) at $\Pi d \approx k_* d$ for $k_* d \gg 1$.
Consequently, a peak structure in $-\text{Exp}[V]$ appears for any value of $k_* d$.
The variance $\text{Var}[V]$ in the case of $k_* d \gg 1$ again shows a richer structure in the $\omega$-dependence (see Fig.~\ref{fig:V});
it remains almost constant in the frequency range $1 \lesssim \Pi d \lesssim k_* d$.
We give an analytic explanation for these behaviors of the Stokes parameters in Sec.~\ref{subsec:analytic}.

As shown in Figs.~\ref{fig:I} and \ref{fig:V}, when the system parameters take our fiducial values, such as $B_{*} \sim \text{nG}$, $k_* \sim \text{Mpc}^{-1}$, $d \sim 100 \,\text{Mpc}$, $n_e \sim \text{m}^{-1}$, the notable transitions in the $\omega$-dependence of the Stokes parameters occur only at extremely high frequencies, $\omega \gtrsim 100 \, \text{eV} \sim 10^{16}\, \text{Hz}$.
Possible sources of high-frequency GWs may include light primordial black holes~\cite{LISACosmologyWorkingGroup:2023njw, Carr:2023tpt}, high-energy phenomena in the early universe (such as inflation 
\cite{Starobinsky:1979ty}, preheating~\cite{Kofman:1994rk}, dynamics of topological defects~\cite{Vilenkin:1981bx,Vachaspati:1984gt}, cosmological phase transitions~\cite{Witten:1984rs,Hogan:1986dsh}, the high-temperature thermal bath~\cite{Ghiglieri:2015nfa,Ghiglieri:2020mhm,Ringwald:2020ist,Vagnozzi:2022qmc}), thermal plasma in stellar interiors~\cite{weinberg2013gravitation}, and photon-to-graviton conversion in a magnetosphere of black holes~\cite{Saito:2021sgq}, and so on (for reviews, see Refs.~\cite{Aggarwal:2020olq,Aggarwal:2025noe}).
Noting that one of the transition points is given by $\Pi (\propto \omega^{-1}) \approx k_*$, a transition at lower frequencies may be realized by a larger $k_*$ (i.e., a shorter correlation length).
While it is experimentally challenging to observe such a frequency dependence, it may provide a potential probe of cosmic magnetic fields in the (far) future.

We conclude by mentioning several possible directions for future work.
First, in the present paper we assumed unpolarized GWs as the initial state, bearing an incoherent superposition of GWs emitted from many sources with no preferred polarization in mind.
If the initial state has a nontrivial polarization, it may affect the final observables.
Second, in this paper we considered a single statistically uniform region characterized by magnetic field power spectra.
However, in more realistic situations, the power spectra are expected to vary across different regions, such as intra-galactic and inter-galactic spaces.
It will be important to develop a formulation that is generally applicable to such cases.
Finally, it will be interesting to apply the framework developed in this paper for graviton-photon conversion to systems involving other light particles. A notable example will be the axion-photon system~\cite{PhysRevD.37.1237}, where the conversion to axions is expected to affect the polarization state of electromagnetic waves.
We leave these studies for future work.

\begin{acknowledgments}

The authors are grateful to Asuka Ito, Kohei Kamada, and Jiro Soda for helpful discussions.
The work of R.J. is supported by JSPS KAKENHI Grant Numbers 23K17687, 23K19048, and 24K07013.
K.N. was supported by JSPS KAKENHI Grant Number JP24KJ0117. 

\end{acknowledgments}

\appendix

\section{Wick contractions of multipoint function}
\label{app:Wick}

Observables associated with GWs are given by the Stokes parameters.
Among them we discuss only $I$ and $V$, since the other two vanish when the initial state is assumed to be unpolarized GWs (see Eq.~\eqref{eq:Stokes_statistical}).
As shown in Eqs.~\eqref{eq:Ia} and \eqref{eq:Vb}, they are expressed as 
\begin{align}
    \begin{cases}
    I = 1 - a \,, \\[0.1cm]
    V = - i \, b \,,
  \end{cases}
\end{align}
where 
\begin{align}
    \begin{cases}
          a = \displaystyle\frac{1}{2M_{\rm P}^2} \int_{s,s'} e^{-i \Pi (s-s')}
          [ B_x(s) B_x(s') + B_y(s) B_y(s') ]
          \,,\\[0.3cm]
          b = \displaystyle\frac{1}{2M_{\rm P}^2} \int_{s,s'} e^{-i \Pi (s-s')}
          [ B_x(s) B_y(s') - B_y(s) B_x(s') ]\,.
    \end{cases}
\end{align}
By taking the ensemble average over the magnetic field, we obtain the statistical averages of these observables.
Since the magnetic field is encoded in $a$ and $b$, we compute the statistical quantities associated with them.
In the following calculation, we use the relation $\hat{k}^2_x=\hat{k}^2_y=(\hat{k}^2_x+\hat{k}^2_y)/2=(1-\hat{k}^2_z)/ 2$ under integration in the wavevector space, which follows from the rotational symmetry in the $x$-$y$ plane.

\subsection{\texorpdfstring{Calculation of $\left< a \right>$ and $\left< b \right>$}{Calculation of <a> and <b>}}

We first calculate the expectation values of the Stokes parameters.
The expectation value of the intensity $\mathrm{Exp}[I]=1-\left< a \right>$ is calculated as
\begin{align}
    \left< a\right>
    &= \frac{1}{2M_{\rm P}^2} \int_{s,s'} e^{-i \Pi (s-s')}
          \left< B_x(s) B_x(s') + B_y(s) B_y(s') \right>
    =\frac{1}{2 M_{\rm P}^2} \int_{s,s'} e^{-i\Pi(s-s')} \int_k e^{ik_z(s-s')} (2-\hat{k}^2_x-\hat{k}^2_y) P_B(k)
    \nonumber\\
    &= \frac{1}{M_{\rm P}^2} \int_k \int_{s,s'} e^{-i(\Pi-k_z)(s-s')} \frac{1+\hat{k}_z^2}{2}P_B(k)
    = \frac{1}{M_{\rm P}^2} \int_k\mathcal{I}(k_z;\omega)\frac{1+\hat{k}^2_z}{2}P_{B}(k)
    \equiv 2\alpha\,.
\end{align}
Here $\mathcal{I}(k_z; \omega)$ is defined as
\begin{align}
  \mathcal{I}(k_z;\omega) 
  \equiv \int_{s,s'} e^{-i(\Pi(\omega)-k_z)(s-s')}
    = \bigg|\int_{s}e^{-i(\Pi(\omega)-k_z)s}\bigg|^2
    =\mathcal{I}^*(k_z;\omega)\,.
  \label{eq:calI}
\end{align}
Consequently, we get
\begin{align}
    \mathrm{Exp}[I]=1-2\alpha\,.
\end{align}
The other quantity is the expectation value of the circular polarization $\mathrm{Exp}[V]=\mathrm{Im}[\left< b\right>]$ calculated as
\begin{align}
    \left< b\right>
    &= \frac{1}{2M_{\rm P}^2} \int_{s,s'} e^{-i \Pi (s-s')}
          \left< B_x(s) B_y(s') - B_y(s) B_x(s') \right>
    = - \frac{i}{M_{\rm P}^2} \int_k \left(\int_{s,s'}e^{-i(\Pi-k_z)(s-s')}\right) \hat{k}_zP_{aB}(k)
    \nonumber\\
    &= - \frac{i}{M_{\rm P}^2} \int_k \mathcal{I}(k_z;\omega) \hat{k}_z P_{aB}(k) 
    \equiv -2i \beta\,.
\end{align}
Thus we get
\begin{align}
    \mathrm{Exp}[V]= -2\beta\,.
\end{align}

\subsection{\texorpdfstring{Calculation of $\left< a^2 \right>$}{Calculation of <a2>}}

We move on to the calculation of variances.
The variance of the intensity $\mathrm{Var}[I]=\left< a^2\right>-\left< a\right>^2$ is calculated as
\begin{align}
   \left< a^2 \right>
   &= \left( \frac{1}{2 M_{\rm P}^2} \right)^2 
   \left<
   \left(\int_{s,s'}e^{-i\Pi(s-s')}[B_x(s)B_x(s')+ B_y(s)B_y(s')]\right)^2
   \right>
   \nonumber\\
   &=\left( \frac{1}{2 M_{\rm P}^2} \right)^2
   \left< 
   \left(\int_{s,s'}e^{-i\Pi (s-s')}B_x(s)B_x(s')\right) \left(\int_{s,s'}e^{-i\Pi (s-s')}B_x(s)B_x(s')\right)
   \right> 
   \nonumber \\
   &~~~~+ \left( \frac{1}{2 M_{\rm P}^2} \right)^2
   \left< 
   \left(\int_{s,s'}e^{-i\Pi (s-s')}B_x(s)B_x(s')\right) \left(\int_{s,s'}e^{-i\Pi (s-s')}B_y(s)B_y(s')\right)
   \right>
   \nonumber\\
   &~~~~+ \left( \frac{1}{2 M_{\rm P}^2} \right)^2
   \left< 
   \left(\int_{s,s'}e^{-i\Pi (s-s')}B_y(s)B_y(s')\right) \left(\int_{s,s'}e^{-i\Pi (s-s')}B_x(s)B_x(s')\right)
   \right> 
   \nonumber\\
   &~~~~+ \left( \frac{1}{2 M_{\rm P}^2} \right)^2
   \left< 
   \left(\int_{s,s'}e^{-i\Pi (s-s')}B_y(s)B_
   y(s')\right) \left(\int_{s,s'}e^{-i\Pi (s-s')}B_y(s)B_y(s')\right)
   \right> 
   \nonumber \\
   &\equiv I_1+I_2+I_3+I_4 \,.
\end{align}
At the level of expectation values, it is evident that the first and fourth terms are identical since there is no statistical difference in the $x$ and $y$ directions, as are the second and third terms.
Thus we can write
\begin{align}
     \left< a^2\right> =2(I_1+I_2)\,.
\end{align}
In the following we adopt the unit $\sqrt{2}M_{\rm P} = 1$ for simplicity.
The first quantity $I_1$ is decomposed as
\begin{align}
    I_1&=\left< \left(\int_{s,s'}e^{-i\Pi (s-s')}B_i(s)B_i(s')\right)^2\right>
    \nonumber \\
    &=\int_{s,s',s'',s'''}e^{-i\Pi (s-s'+s''-s''')}\left< B_i(s)B_i(s')B_i(s'')B_i(s''')\right>\nonumber\\
    &=\int_{s,s',s'',s'''}e^{-i\Pi (s-s'+s''-s''')}
    \nonumber \\
    &~~~~
    \times\left[\left< B_i(s)B_i(s')\right>\left<  B_i(s'')B_i(s''')\right>+\left< B_i(s)B_i(s'')\right>\left<  B_i(s')B_i(s''')\right>+\left< B_i(s)B_i(s''')\right>\left<  B_i(s')B_i(s'')\right>\right]\nonumber\\
    &\equiv A_1+A_2+A_3
    \,.
\end{align}
The first term is
\begin{align}
    A_1&=\int_{s,s',s'',s'''}e^{-i\Pi (s-s'+s''-s''')}\left< B_i(s)B_i(s')\right>\left<  B_i(s'')B_i(s''')\right>\nonumber\\
    &=\int_{k,k'}\int_{s,s',s'',s'''}e^{-i\Pi (s-s'+s''-s''')}e^{ik_z (s-s')}e^{ik'_z (s''-s''')}(1-\hat{k}_i^2)(1-\hat{k}'^2_i)P_B(k)P_B(k')\nonumber\\
    &=\int_{k,k'}\mathcal{I}(k_z;\omega) \, \mathcal{I}(k'_z;\omega)\left(\frac{1+\hat{k}^2_z}{2}\right)\left(\frac{1+\hat{k}'^2_z}{2}\right)P_B(k)P_B(k')\nonumber\\
    &=\left[\int_k\mathcal{I}(k_z;\omega)\left(\frac{1+\hat{k}^2_z}{2}\right)P_B(k)\right]^2\nonumber\\
    &\equiv
    \alpha^2\,.
\end{align}
Here $\mathcal{I}(k_y; \omega)$ is defined in Eq.~\eqref{eq:calI}.
The second term is
\begin{align}
    A_2&=\int_{s,s',s'',s'''}e^{-i\Pi (s-s'+s''-s''')}\left< B_i(s)B_i(s'')\right>\left<  B_i(s')B_i(s''')\right>\nonumber\\
    &=\int_{k,k'}\int_{s,s',s'',s'''}e^{-i\Pi (s-s'+s''-s''')}e^{ik_z (s-s'')}e^{ik'_z (s'-s''')}(1-\hat{k}_i^2)(1-\hat{k}'^2_i)P_B(k)P_B(k')\nonumber\\
    &=\int_{k,k'}\bar{\mathcal{J}}(k_z;\omega)\bar{\mathcal{J}}^*(k'_z;\omega)\left(\frac{1+\hat{k}^2_z}{2}\right)\left(\frac{1+\hat{k}'^2_z}{2}\right)P_B(k)P_B(k')\nonumber\\
     &=\left[\int_k\mathcal{J}(k_z;\omega)\left(\frac{1+\hat{k}^2_z}{2}\right)P_B(k)\right]^2\nonumber\\
     &\equiv
     \gamma^2\,.
\end{align}
Here $\bar{\mathcal{J}}(k_z;\omega)$ is defined as 
\begin{align}
    \bar{\mathcal{J}}(k_z;\omega)\equiv \int_s e^{-i(\Pi(\omega) - k_z)s}\int_{s'}e^{-i(\Pi(\omega) + k_z)s'}=\bar{\mathcal{J}}(-k_z;\omega)\,,
\end{align}
and $\mathcal{J}$ is defined as the absolute value of $ \bar{\mathcal{J}}$.
The last term is
\begin{align}
    A_3&=\int_{s,s',s'',s'''}e^{-i\Pi (s-s'+s''-s''')}\left< B_i(s)B_i(s''')\right>\left<  B_i(s')B_i(s'')\right>
    = A_1 = \alpha^2\,.
\end{align}
We thus obtain the first term of $\left< a^2\right>$ as
\begin{align}
    I_1= 2\alpha^2+\gamma^2 \,.
\end{align}
The second quantity $I_2$ involves magnetic field components of different orientations.
It is decomposed as
\begin{align}
    I_2&=\left< \left(\int_{s,s'}e^{-i\Pi (s-s')}B_x(s)B_x(s')\right)~\times\left(\int_{s,s'}e^{-i\Pi (s-s')}B_y(s)B_y(s')\right)\right> 
    \nonumber \\
    &=\int_{s,s',s'',s'''}e^{-i\Pi (s-s'+s''-s''')}
    \left< B_x(s) B_x(s') B_y(s'') B_y(s''') \right>\nonumber\\
    &=\int_{s,s',s'',s'''}e^{-i\Pi (s-s'+s''-s''')}
    \nonumber \\
    &~~~~
    \times\left[\left< B_x(s)B_x(s')\right>\left<  B_y(s'')B_y(s''')\right>+\left< B_x(s)B_y(s'')\right>\left<  B_x(s')B_y(s''')\right>+\left< B_x(s)B_y(s''')\right>\left<  B_x(s')B_y(s'')\right>\right]\nonumber\\
    &\equiv B_1+B_2+B_3
    \,.
\end{align}
The first term is
\begin{align}
    B_1&=\int_{s,s',s'',s'''}e^{-i\Pi (s-s'+s''-s''')}\left< B_x(s)B_x(s')\right> \left<  B_y(s'')B_y(s''')\right>
    \nonumber\\
    &=\int_{k,k'}\int_{s,s',s'',s'''}e^{-i\Pi (s-s'+s''-s''')}e^{ik_z (s-s')}e^{ik'_z (s''-s''')}(1-\hat{k}_x^2)(1-\hat{k}'^2_y)P_B(k)P_B(k')\nonumber\\
    &=\int_{k,k'}\mathcal{I}(k_z;\omega) \, \mathcal{I}(k'_z;\omega)\left(\frac{1+\hat{k}^2_z}{2}\right)\left(\frac{1+\hat{k}'^2_z}{2}\right)P_B(k)P_B(k')\nonumber\\
    &=\left[\int_k\mathcal{I}(k_z;\omega)\left(\frac{1+\hat{k}^2_z}{2}\right)P_B(k)\right]^2\nonumber\\
    &\equiv
    \alpha^2\,.
\end{align}
The second term is
\begin{align}
    B_2&=\int_{s,s',s'',s'''}e^{-i\Pi (s-s'+s''-s''')}\left< B_x(s)B_y(s'')\right>\left<  B_x(s')B_y(s''')\right>\nonumber\\
    &=\int_{k,k'}\int_{s,s',s'',s'''}e^{-i\Pi (s-s'+s''-s''')}e^{ik_z (s-s'')}e^{ik'_z (s'-s''')}(-i\hat{k}_z)(-i\hat{k}_z')P_{aB}(k)P_{aB}(k')\nonumber\\
    &=-\int_{k,k'}\bar{\mathcal{J}}(k_z;\omega)\bar{\mathcal{J}}^*(k'_z;\omega)\hat{k}_z\hat{k}_z'P_{aB}(k)P_{aB}(k')\nonumber\\
    &=-\left[\int_k\mathcal{J}(k_z;\omega)\hat{k}_z P_{aB}(k)\right]^2\nonumber\\
    &\equiv
    -\delta^2\,.
\end{align}
However, because of the property $\bar{\mathcal{J}}(k_z;\omega) = \bar{\mathcal{J}}(-k_z;\omega)$, this term is identically vanishing
\begin{align}
    \delta = 0\,. 
\end{align}
The last term is
\begin{align}
    B_3&=\int_{s,s',s'',s'''}e^{-i\Pi (s-s'+s''-s''')}\left< B_x(s)B_y(s''')\right>\left<  B_x(s')B_y(s'')\right>\nonumber\\
    &=\int_{k,k'}\int_{s,s',s'',s'''}e^{-i\Pi (s-s'+s''-s''')}e^{ik_z (s-s''')}e^{ik'_z (s'-s'')}(-i\hat{k}_z)(-i\hat{k}_z')P_{aB}(k)P_{aB}(k')\nonumber\\
    &=-\int_{k,k'}\mathcal{I}(k_z;\omega)\mathcal{I}(-k'_z;\omega)\hat{k}_z\hat{k}'_zP_{aB}(k)P_{aB}(k')\nonumber\\
    &=\left[\int_k\mathcal{I}(k_z;\omega)\hat{k}_z P_{aB}(k)\right]^2\nonumber\\
    &\equiv
    \beta^2\,.
\end{align}
Thus the second quantity $I_2$ is
\begin{align}
    I_2=\alpha^2-\delta^2+\beta^2\,.
\end{align}
To summarize, the statistical variance of the intensity becomes
\begin{align}
    \mathrm{Var}[I]=\left< a^2\right>-\left< a\right>^2
    &=2(\alpha^2+\beta^2+\gamma^2-\delta^2)\,.
\end{align}
Through the preceding calculations, one may notice several recurring structures.
These are classified by (1) whether the magnetic field components in the two-point function are the same or different, and (2) whether the coordinate integrals give the kernel $\mathcal{I}$ or $\mathcal{J}$.

\subsection{\texorpdfstring{Calculation of $\left< b^2\right>$}{Calculation of <b2>}}

The variance of the Stokes parameter $\mathrm{Var}[V]=-(\left< b^2\right>-\left< b\right>^2)$ is calculated as
\begin{align}
   b^2&=\left( \frac{1}{2 M_{\rm P}^2} \right)^2 \left(\int_{s,s'}e^{-i\Pi(s-s')}[B_x(s)B_y(s')- B_y(s)B_x(s')]\right)^2\nonumber\\
   &=\left( \frac{1}{2 M_{\rm P}^2} \right)^2 \left(\int_{s,s'}e^{-i\Pi (s-s')}B_x(s)B_y(s')\right)\left(\int_{s,s'}e^{-i\Pi (s-s')}B_x(s)B_y(s')\right)\nonumber\\
   &\quad-\left( \frac{1}{2 M_{\rm P}^2} \right)^2 \left(\int_{s,s'}e^{-i\Pi (s-s')}B_x(s)B_y(s')\right)\left(\int_{s,s'}e^{-i\Pi (s-s')}B_y(s)B_x(s')\right)\nonumber\\
   &\quad-\left( \frac{1}{2 M_{\rm P}^2} \right)^2 \left(\int_{s,s'}e^{-i\Pi (s-s')}B_y(s)B_x(s')\right)\left(\int_{s,s'}e^{-i\Pi (s-s')}B_x(s)B_y(s')\right)\nonumber\\
   &\quad+\left( \frac{1}{2 M_{\rm P}^2} \right)^2 \left(\int_{s,s'}e^{-i\Pi (s-s')}B_y(s)B_
   x(s')\right)\left(\int_{s,s'}e^{-i\Pi (s-s')}B_y(s)B_x(s')\right)
   \nonumber \\
   &\equiv J_1 - J_2 - J_3 + J_4 \,.
\end{align}
At the level of expectation values, it is evident that the first and fourth terms are identical, as are the second and third terms:
\begin{align}
     \left< b^2\right>\equiv J_1-J_2-J_3+J_4=2(J_1-J_2)\,.
\end{align}
Similarly to the previous subsection, we adopt the unit $\sqrt{2}M_{\rm P} = 1$ for simplicity.
By systematically performing the contractions and paying attention to the magnetic field components and integral functions, the calculation proceeds in a straightforward way.
We obtain the first term of $\left< b^2\right>$ as
\begin{align}
    J_1&= \left< \left(\int_{s,s'}e^{-i\Pi (s-s')}B_x(s)B_y(s')\right)~\left(\int_{s,s'}e^{-i\Pi (s-s')}B_x(s)B_y(s')\right)\right>=-\beta^2+\gamma^2-\beta^2\,.
\end{align}
Similarly, we obtain the second term of $\left< b^2\right>$ as
\begin{align}
    J_2&= \left< \left(\int_{s,s'}e^{-i\Pi (s-s')}B_x(s)B_y(s')\right)~\left(\int_{s,s'}e^{-i\Pi (s-s')}B_y(s)B_x(s')\right)\right>=\beta^2+\delta^2+\alpha^2\,.
\end{align}
To summarize, the variance of Stokes parameter $V$ is
\begin{align}
    \mathrm{Var}[V]
    =-(\left< b^2\right>-\left< b\right>^2)
    =-2[(-2\beta^2+\gamma^2)-(\beta^2+\delta^2+\alpha^2)]-(2\beta)^2
    =2(\alpha^2+\beta^2-\gamma^2+\delta^2)\,.
\end{align}

\section{Expressions for the convolution kernels}
\label{app:C}

In this appendix we summarize the expressions for the convolution kernels in Eq.~\eqref{eq:C}:
\begin{align}
\mathcal{C}_\alpha (x = \Pi d, y = kd)
&=
- \frac{1}{y (x - y z)}
\left[
2 x (x-y z) (\text{Ci}(| x-y z| ) - \ln (| x-y z| ))+\left(x^2+y^2\right) (x-y z) \text{Si}(x-y z)
\right.
\nonumber \\
&\qquad\qquad\qquad\quad
\left.
+(x^2+y^2) \cos (x-y z)-(x-y z) \sin (x-y z)-x^2-xy z-y^2+y^2 z^2
\right]
\Big|_{z = -1}^{z = 1}
\,,
\\
\mathcal{C}_\beta (x = \Pi d, y = kd)
&=
-\frac{2}{x-y z}
\left[
(x-y z) (\text{Ci}(| x-y z| )-\ln (| x-y z| ))+x (x-y z) \text{Si}(x-y z)+x \cos (x-y z)-x
\right]
\Big|_{z = -1}^{z = 1}
\,,
\\
\mathcal{C}_\gamma (x = \Pi d, y = kd)
&=
\frac{1}{2 x y}
\left[
\left(x^2+y^2\right) \cos x (\text{Ci}(| x+y z| )-\ln (| x+y z| )-\text{Ci}(| x-y z| )  + \ln (| x-y z| ))
\right.
\nonumber \\
&\qquad\qquad
\left.
+\left(x^2+y^2\right) \sin x (\text{Si}(x+y z)-\text{Si}(x-y z))
+2 x (y z \cos x-\sin (y z))
\right]
\Big|_{z = -1}^{z = 1}
\,.
\end{align}

\section{Non-helical case}
\label{app:non-helical}

\begin{figure}[H]
\centering
\includegraphics[width=0.45\linewidth]{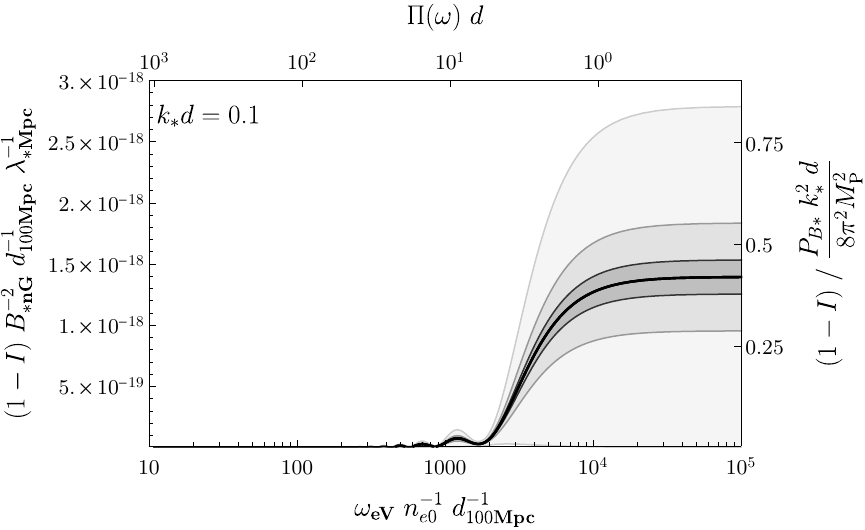}
\hskip 0.5cm
\includegraphics[width=0.45\linewidth]{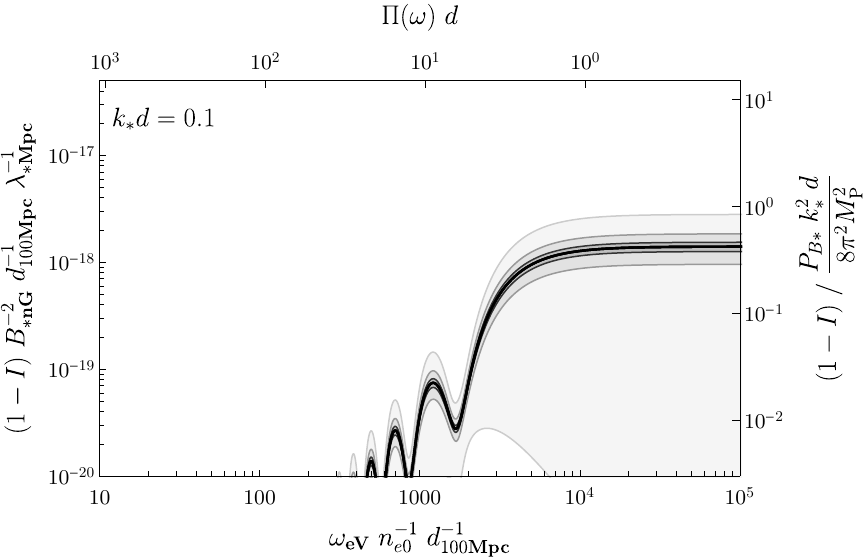}
\\
\includegraphics[width=0.45\linewidth]{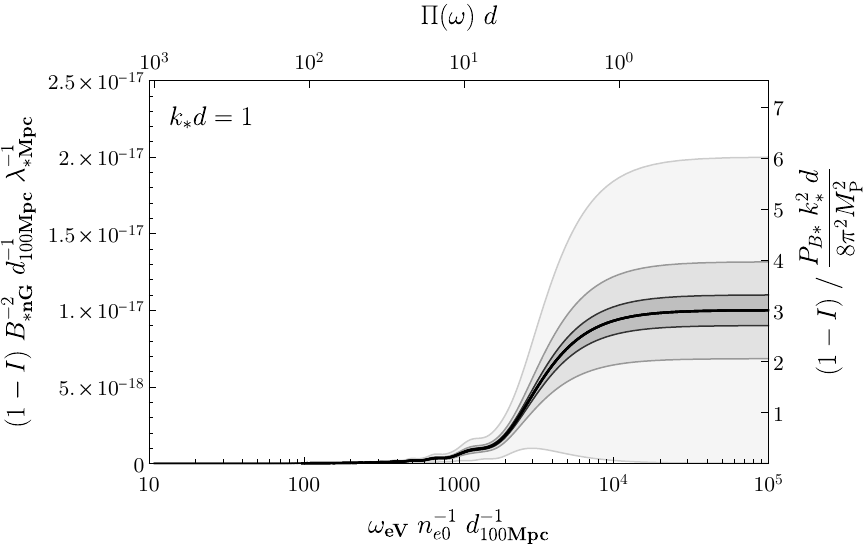}
\hskip 0.5cm
\includegraphics[width=0.45\linewidth]{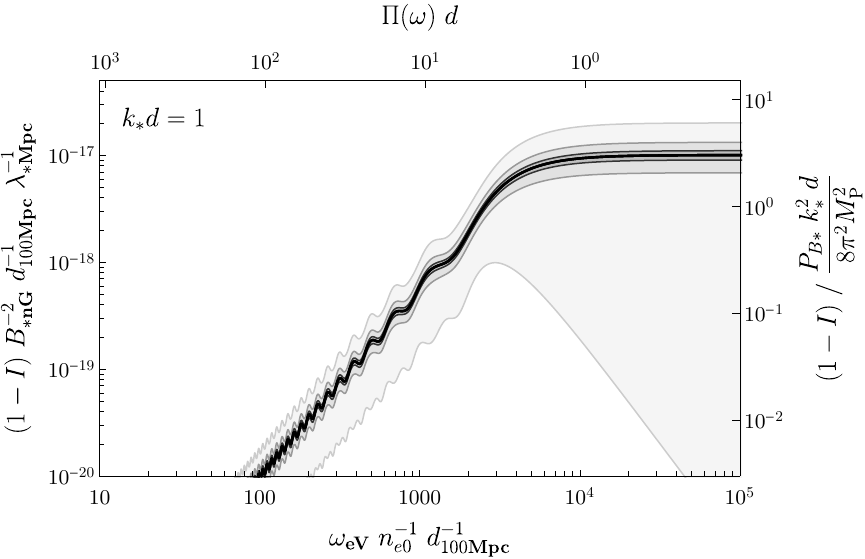}
\\
\includegraphics[width=0.45\linewidth]{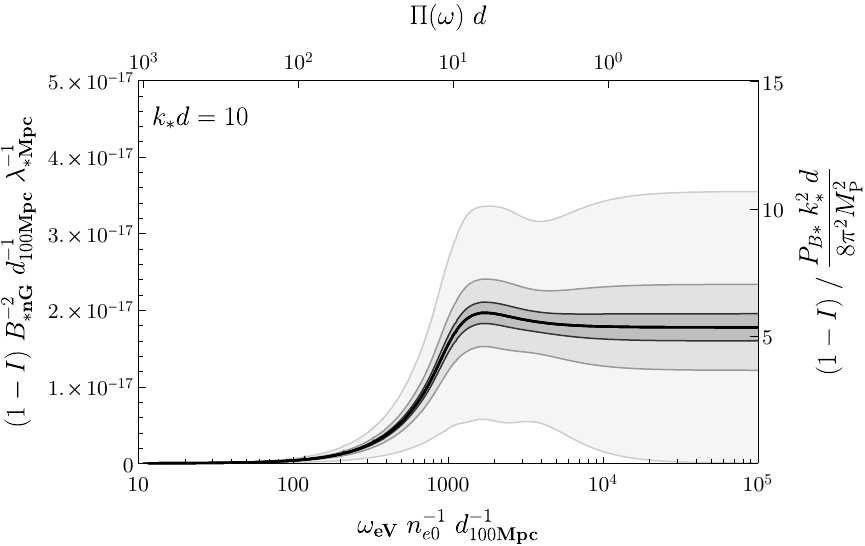}
\hskip 0.5cm
\includegraphics[width=0.45\linewidth]{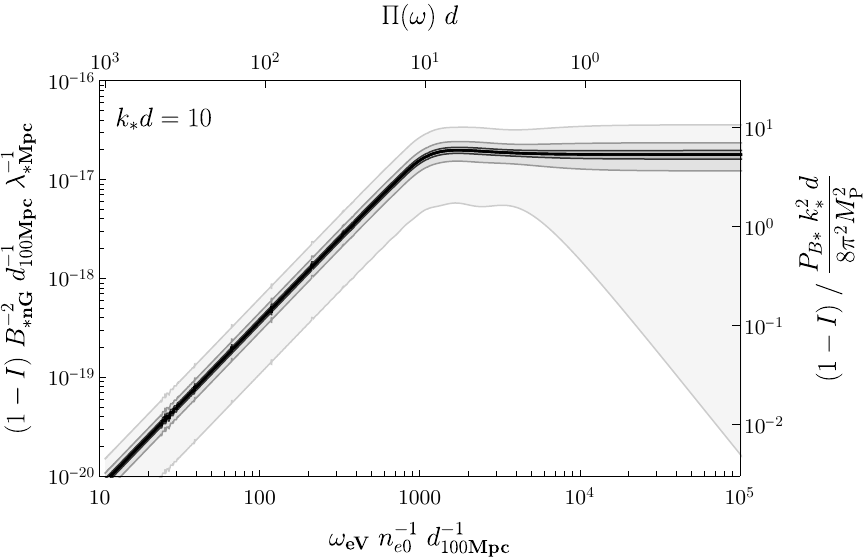}
\\
\includegraphics[width=0.45\linewidth]{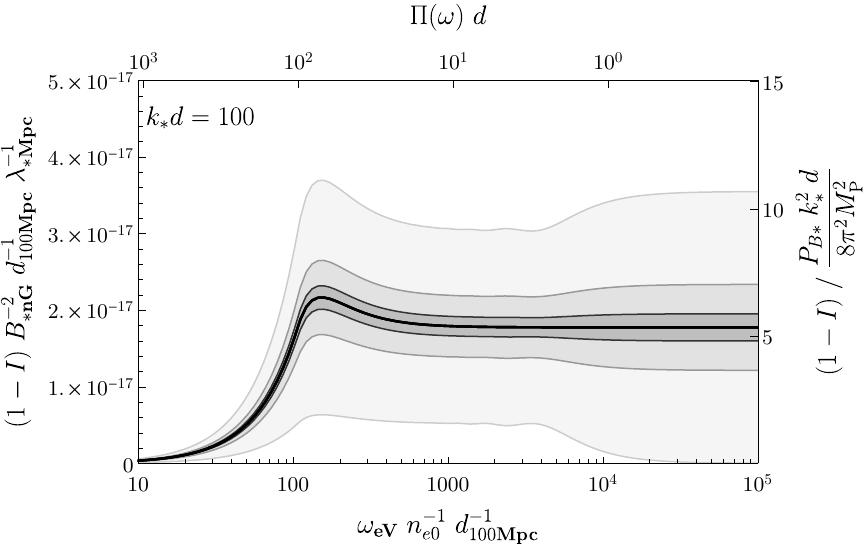}
\hskip 0.5cm
\includegraphics[width=0.45\linewidth]{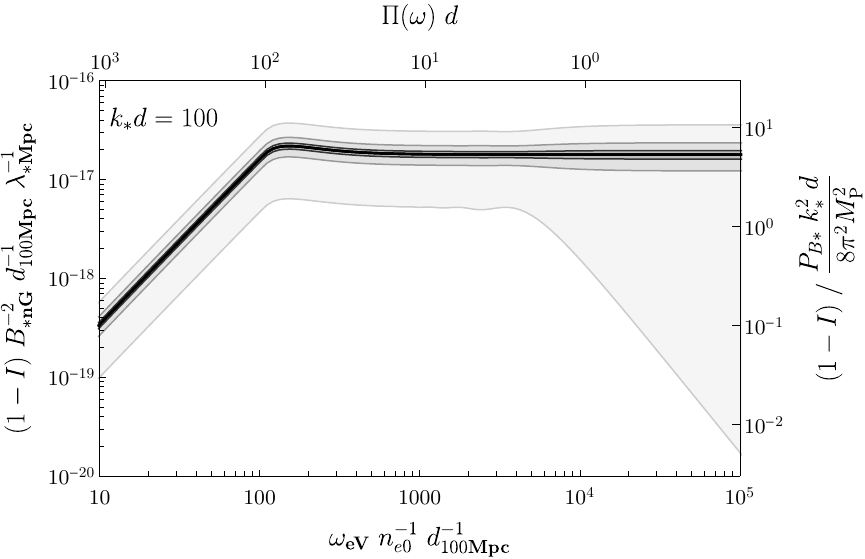}
\caption{\small
Conversion probability $P_{g\rightarrow\gamma} = 1-I$ of GWs with the non-helical ($P_{aB*} =0$) magnetic field power spectrum \eqref{eq:PB} is shown as a function of GW angular frequency $\omega$ for various values of $k_* d$.
From top to bottom, the cases $k_* d = 0.1$, $1$, $10$, and $100$ are shown. 
In each panel, the black line represents the expectation value $\text{Exp}[1-I]$, and the shaded bands represent the standard deviations $\sqrt{\text{Var}[1-I]}$, $\sqrt{\text{Var}[1-I]/10}$, and $\sqrt{\text{Var}[1-I]/100}$.
The upper horizontal axis is the dimensionless variable $\Pi(\omega) d$, and the lower horizontal axis is $\omega_{\text{eV}} \equiv \omega / \text{eV}$ normalized by $n_{e0} \equiv n_e / \text{m}^{-3}$ and $d_{100\text{Mpc}} \equiv d / (100\,\text{Mpc})$, where $n_e$ is the plasma density and $d$ is the propagation distance. 
The right vertical axis is $1-I$ normalized by $P_{B*} k_*^2 d / (8\pi^2 M_{\rm P}^2)$. 
The left vertical axis is $1-I$ normalized by $B_{*\text{nG}}^2 \equiv (B_* / \text{nG})^2$, $d_{100\text{Mpc}}$, and $\lambda_{*\text{Mpc}} \equiv \lambda_{*} / \text{Mpc}$, where $B_{*}$ is the magnetic field strength at the peak scale $k_*$, and $\lambda_{*} \equiv 2\pi/k_*$ is the characteristic length.
In each row, the left and right panels show the same quantity: The vertical axis is linear in the left, but logarithmic in the right.
}
\label{fig:PlotInon}
\end{figure}

\begin{figure}[H]
\centering
\includegraphics[width=0.45\linewidth]{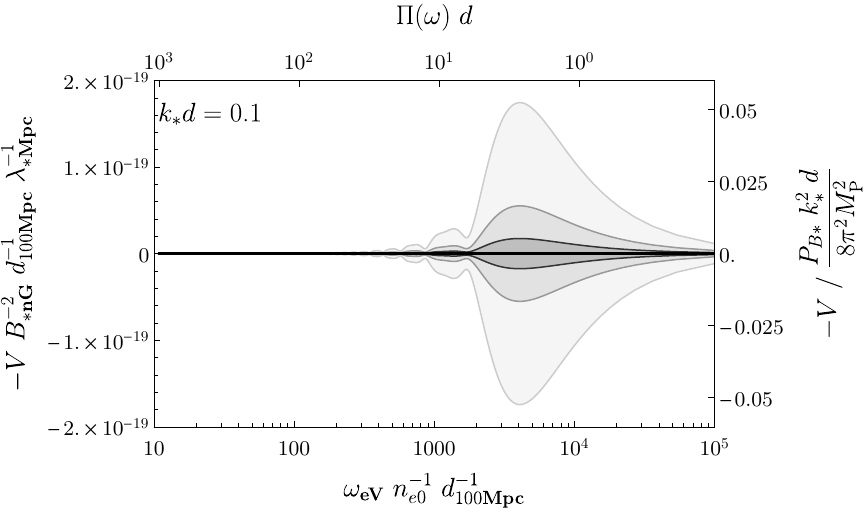}
\hskip 0.5cm
\includegraphics[width=0.45\linewidth]{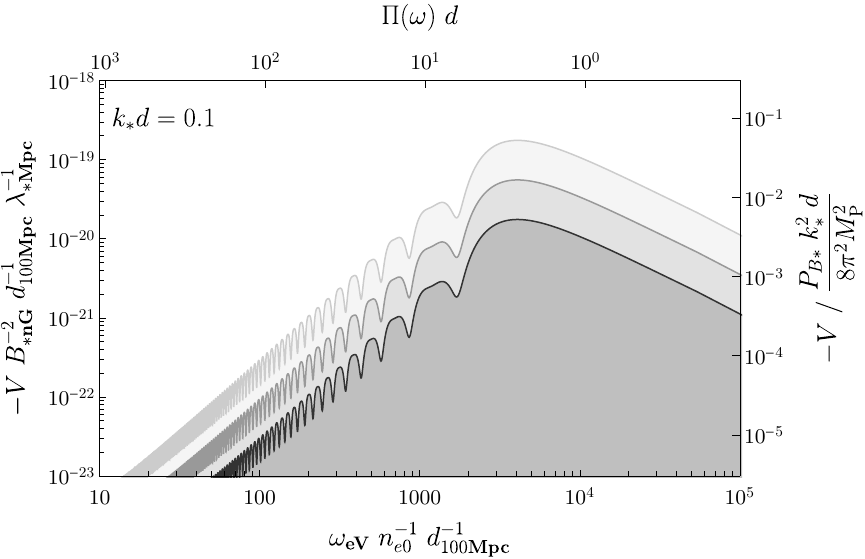}
\\
\includegraphics[width=0.45\linewidth]{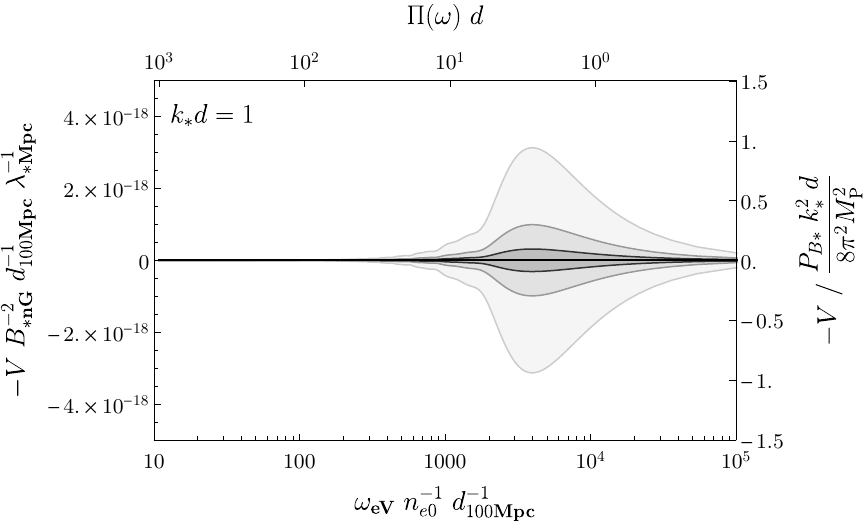}
\hskip 0.5cm
\includegraphics[width=0.45\linewidth]{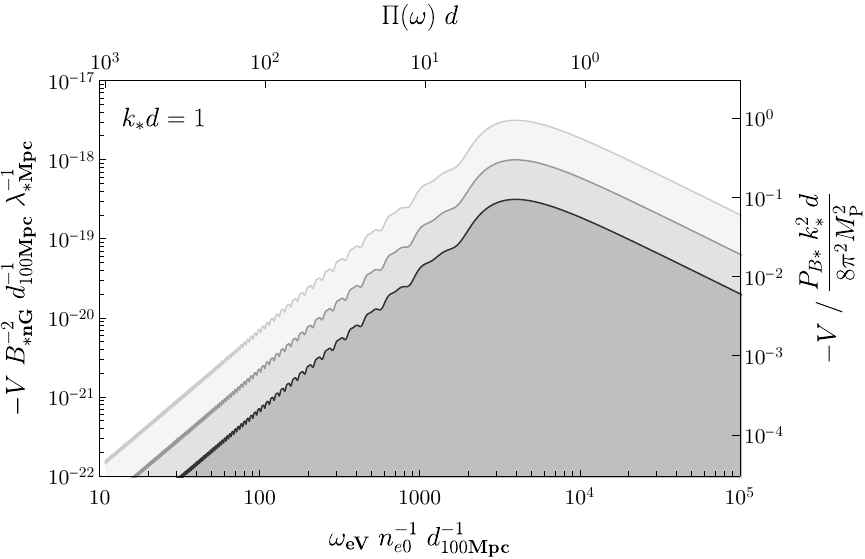}
\\
\includegraphics[width=0.45\linewidth]{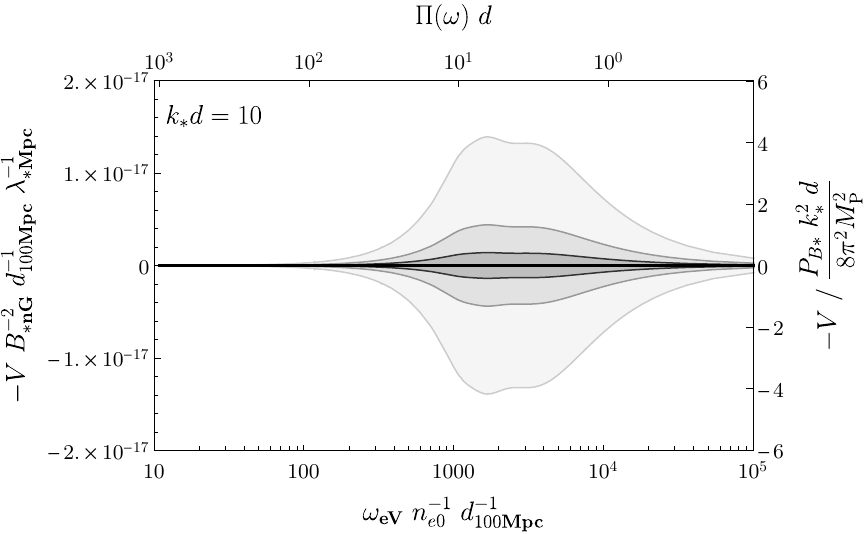}
\hskip 0.5cm
\includegraphics[width=0.45\linewidth]{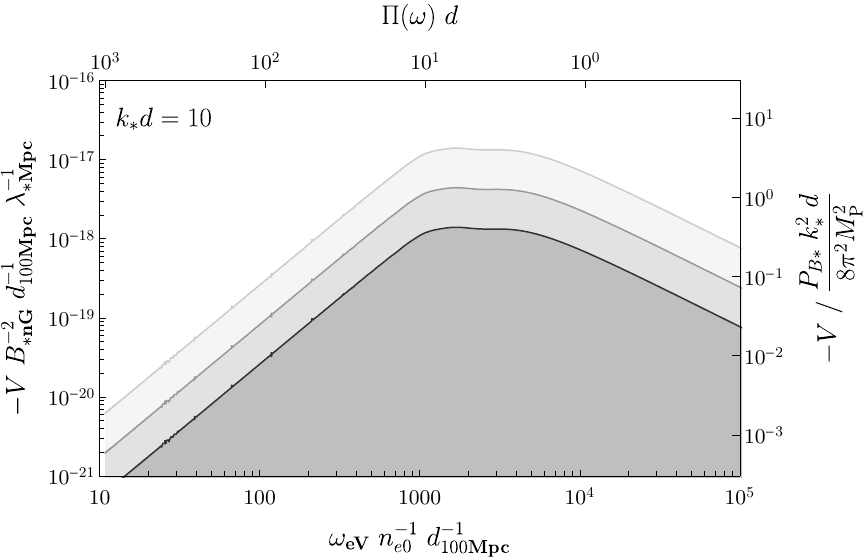}
\\
\includegraphics[width=0.45\linewidth]{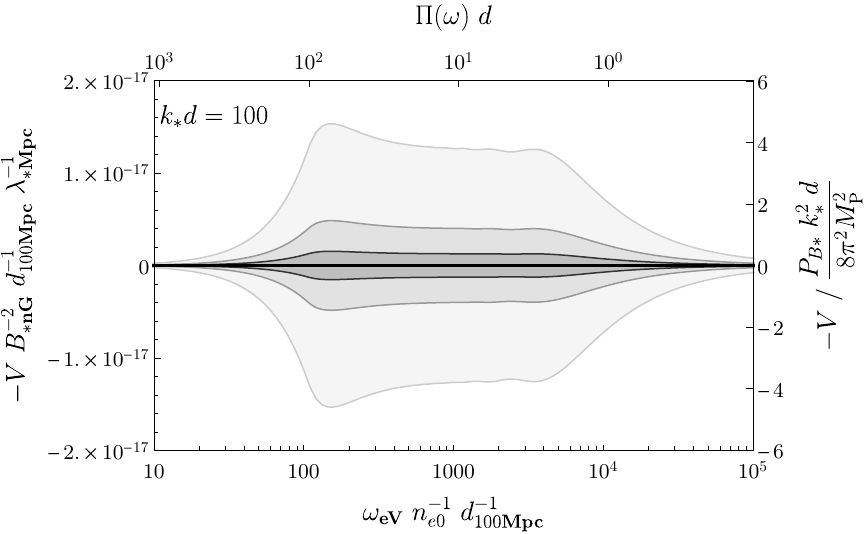}
\hskip 0.5cm
\includegraphics[width=0.45\linewidth]{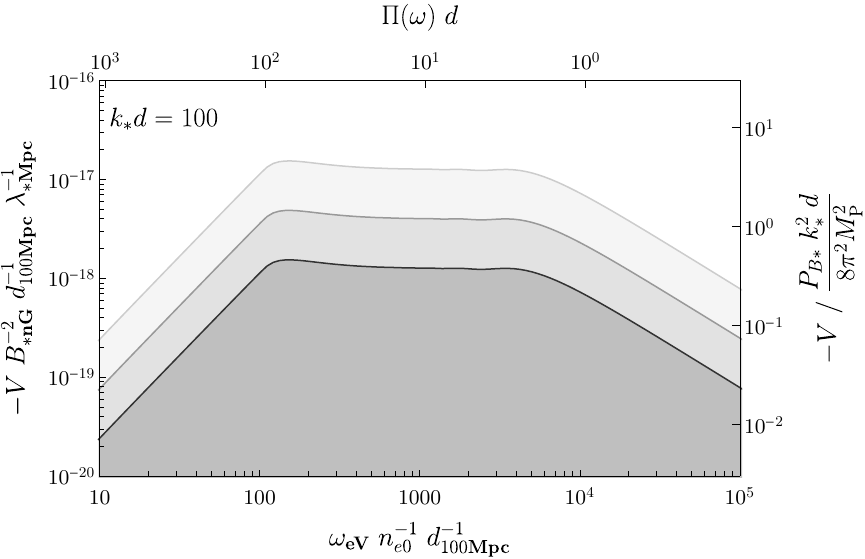}
\caption{\small
Stokes parameter $-V$ (circular polarization) of GWs with the non-helical ($P_{aB*} =0$) magnetic field power spectrum \eqref{eq:PB} is shown as a function of the angular frequency $\omega$ for various values of $k_* d$.
From top to bottom, the cases $k_* d = 0.1$, $1$, $10$, and $100$ are shown.
In each panel, the black line represents the expectation value $-\text{Exp}[V]$, and the shaded bands represent the standard deviations $\sqrt{\text{Var}[V]}$, $\sqrt{\text{Var}[V]/10}$, and $\sqrt{\text{Var}[V]/100}$.
The upper horizontal axis is the dimensionless variable $\Pi (\omega) d$, and the lower horizontal axis is $\omega_{\text{eV}} \equiv \omega / \text{eV}$ normalized by $n_{e0} \equiv n_e / \text{m}^{-3}$ and $d_{100\text{Mpc}} \equiv d / (100\,\text{Mpc})$, where $n_e$ is the plasma density and $d$ is the propagation distance. 
The right vertical axis is $-V$ normalized by $P_{B*} k_*^2 d / (8\pi^2 M_{\rm P}^2)$. 
The left vertical axis is $-V$ normalized by $B_{*\text{nG}}^2 \equiv (B_* / \text{nG})^2$, $d_{100\text{Mpc}}$, and $\lambda_{*\text{Mpc}} \equiv \lambda_{*} / \text{Mpc}$, where $B_{*}$ is the magnetic field strength at the peak scale $k_*$, and $\lambda_{*} \equiv 2\pi/k_*$ is the characteristic length.
In each row, the left and right panels show the same quantity: The vertical axis is linear in the left, but logarithmic in the right.
}
\label{fig:PlotVnon}
\end{figure}

In this Appendix, we show numerical results for the case of a non-helical magnetic field $P_{aB}=0$.
In Figs.~\ref{fig:PlotInon} and \ref{fig:PlotVnon} we present the behavior of the Stokes parameters $I$ and $V$ as functions of $\Pi(\omega) d$ for various values of $k_* d$, similarly to Figs.~\ref{fig:I} and \ref{fig:V}.
The black lines represent the expectation values while the gray bands correspond to the standard deviations $\sqrt{\mathrm{Var}[1-I]}$ and $\sqrt{\mathrm{Var}[V]}$.
The three different bands are for $(1, 1/\sqrt{10}, 1/\sqrt{100})$ times the original one, respectively.

One observes that Fig.~\ref{fig:PlotInon} remains almost unchanged from Fig.~\ref{fig:I}.
This is because both the expectation value $\text{Exp} [1 - I]$ and the variance $\text{Var} [1 - I]$ are dominated by the integrals $\alpha$ and $\gamma$, which are determined by the power spectrum $P_B$, not by $P_{a B}$.
In contrast, Fig.~\ref{fig:PlotVnon} behaves quite differently from Fig.~\ref{fig:V}.
The main difference is in the expectation value, which vanishes in Fig.~\ref{fig:PlotVnon} since $\text{Exp} [V]$ is determined solely by $\beta$.
On the other hand, the variance $\text{Var} [V]$ remains nonzero because it is largely determined by $\alpha$ and $\gamma$.

\clearpage

\bibliography{main}

\end{document}